# A Survey of fault models and fault tolerance methods for 2D bus-based multi-core systems and TSV based 3D NOC many-core systems.

Shashikiran Venkatesha, Assistant Professor Senior Grade 1, Vellore Institute of Technology Vellore 632014 Tamil Nadu India.

Ranjani Parthasarathi, Professor, College of Engineering Guindy Anna University Chennai 600025 Tamil Nadu India

Corresponding author: Shashikiran Venkatesha            Email: shashikiran.annauniv@gmail.com

Co-author: Ranjani Parthasarathi            Email: ranjani.parthasarathi@gmail.com

**ABSTRACT**

**Reliability has taken centre stage in the development of high-performance computing processors. A Surge of interest is noticeable in recent times in formulating fault and failure models, understanding failure mechanism and strategizing fault mitigation methods for improving the reliability of the system. The article presents a congregation of concepts illustrated one after the other for a better understanding of damages caused by radiation, relevant fault models, and effects of faults. We examine the state of art fault mitigation techniques at the logical layer for digital CMOS based design and SRAM based FPGA. CMOS SRAM structure is the same for both digital CMOS and FPGA. Understanding of resilient SRAM based FPGA is necessary for developing resilient prototypes and it facilitates a faster integration of digital CMOS designs. At the micro-architectural and architectural layer, error detection and recovery methods are discussed for bus-based multi-core systems. The Through silicon via based 3D Network on chip is the prospective solution for integrating many cores on single die. A suitable interconnection approach for petascale computing on many-core systems. The article presents an elaborate discussion on fault models, failure mechanisms, resilient 3D routers, defect tolerance methods for the TSV based 3D NOC many-core systems. Core redundancy, self-diagnosis and distributed diagnosis at the hardware level are examined for many-core systems. The article presents a gamut of fault tolerance solutions from logic level to processor core level in a multi-core and many-core scenario.**

# 1 INTRODUCTION

As stated in the Moore's Law, on chip transistors double every 18 months. The ascendance of the performance of a single-core has followed the foot-steps as stated by Moore's Law [1,2]. The 35% increase in the transistor density every year results in an increase of the die size by 10% to 20% per year. The device speed scales slowly and is governed by Pollack's Rule, which states that performance increases as the square root of increase in complexity [3]. As stated in the Dennard law [4], the circuit latency is enhanced and power overhead is reduced by decreasing the node size of a transistor. It is substantiated by the fact that Intel i386 has an operating frequency at 16Mhz when compared to superscalar processor such as Pentium 4 has an operating frequency of 1.5Ghz. Further reduction in the node size beyond 65nm, ignorance of leakage current and threshold voltage led to failure of Dennard Law. In the year 2006, rapid fall in the device size led to increase power density and created a "power wall" that restricted the operating frequency of a processor at 4 Ghz. The techniques that reduced the clock cycle per instruction for single-core system are Instruction level parallelism (ILP) techniques such as dynamic branch prediction, speculative execution, dynamic scheduling (scoreboard), with renaming (Tomasulo's approach), and multiple instruction issue approaches. Tullsen *et al.* proposed Simultaneous multithreading (SMT) [5], is a technique that process one or more hardware threads concurrently by dividing the resources amongst threads dynamically (e.g., register file, instruction queue, etc.), process known as HyperThreading [6]. The Intel Pentium 4's version of SMT were first saleable SMT processors available in the market that allowed up to 2 hardware threads at a time. Gains in single- core processors are well known, reduction in performance per watt /per mm die area made the architects to re-think and re-strategizes computing resources that led to fabrication of multiple cores on a chip with improved performance and power overheads.

**Need for many-core: -** The multi-core architectures from Intel, AMD, Sun, Nvidia, Tilera, Xilinx and IBM replicate single processor cores many times on a chip die. The commercially available multi-core processors with billions of transistors are listed in the **Table 1.** According to [7], integrating several processing cores does not provide performance better than highest performing single core processor, unless the parallel portion of an application are fully utilized to achieve high performance with low energy cost. The bus-based multi-core architecture were not scalable. The issues in bus-based multi-core architecture are (a) bandwidth is shared and limited (b) timing (c) wire speed is the latency (d) arbitration and (e) testability.

**Table 1 Commercially available multi-core processors**

| Processor | Transistor Count(billions) | Die area (mm$^2$) | Core count | CMOS technology (nm) |
|---|---|---|---|---|
| SPARC T5 | 1.5 | NA | 16 | 28 |
| Xeon Ivy town | 4.31 | NA | 15 | 22 |
| Power 8 | 4.2 | 649 | 12 | 22 |
| IBM System x | 4.0 | 678 | 8 | 22 |
| Xeon E5-2600 | 5.56 | 663.52 | 18 | 22 |
| IBM z13 | 4.0 | 678 | 8 | 22 |
| Qualcomm Snapdragon 835 | 3.0 | 72.3 | 8 | 10 |
| Ryzen 5 PRO 1600 AMD | 4.8 | 213 | 6 | 14 |
| Qualcomm Snapdragon 850 | 5.3 | 94 | 8 | 10 |
| APPLE A12X BIONIC | 10.01 | 122 | 8 | 7 |

**Need for Network on Chip (NOC) based many-cores**: - the problem of bus and global wire delays [8] is mitigated by NOC packet in an on-chip interconnection network by providing provisional common bus with sectional and elastic interconnect formation consisting of small wires. This led to increase in bandwidth and current communication on chip [9,10].

**SRAM based FPGA**: - There are different methods of FPGA programming, they are antifuses, SRAM and EEPROM/Flash. Currently, Static Random-Access Memory (SRAM) based Field Programmable Gate Array (FPGA) are more popular with higher number of reconfigurations cycles supported by them. The gap between the FPGA and Application specific integrated circuits (ASIC) have been reduced for majority of application. Space and Military market demand high performance with flexibility as similar to commercial markets. "Will FPGA kill ASIC" panel in Design Automation conference (2001) have discussed about next generation trends of FPGA. For chip designers, understanding of fault mitigation techniques for SRAM based FPGA will shorten the integration time of building design using digital Complementary Metal Oxides Semiconductors (CMOS). CMOS SRAM structure is same in digital CMOS and SRAM based FPGA. Failure time estimation in SRAM based FPGA are different from the approaches that are adopted in digital CMOS. SRAM based FPGA would be next natural choice for heterogeneous integration in 3D IC designs.

**2D IC designs to Through silicon vias (TSV) based 3D NOC many-core systems**: - TSV based NOC offers performance enhancements [225], higher density of cores and allows integration of different technologies [226] and smaller foot area [224] than traditional 2D IC designs.

**Reliability**: - Rapid advances in CMOS Technology resulted in faster integration of devices and too many challenges. Challenges such as increased power dissipation, thermal dissipation, occurrence of faults in the circuits and reliability issues. Further, faults can be classified in to permanent faults and non-permanent faults. Permanent faults are revealed from process variations and manufacturing defects which reduce processor yield [11]. Environmental condition hazards and races in timing paths cause non-permanent faults which are present only part of time and occur randomly. Transient faults and intermittent faults are two types of non-permanent faults. Radiation, electromagnetic interference and ground loops cause transient faults or soft errors. Understanding the effects of radiation will assist in standardizing logical fault models and failure mechanisms. Researchers expect increase in soft-error rate by 8% for every logic state bit for each technology generation [11]. The failure rate will be 100 times more at 16nm than at 180nm. For many-core systems, CMOS scaling has enabled unprecedented levels of integration, with the downside being a penalty in the power dissipation and reliability. Next to power dissipation, transient and permanent faults are evolving as the upcoming significant challenge for architects designing infallible computer systems.

**Liability of ignoring reliability**: - In the year 1994, errors in floating point divisions [263] unit of Intel Pentium microprocessor resulted in 475 million dollars loss to replace the faulty processors. The particles flipping multiple cache tags arrays [264] of a CPU farm crashed regularly in a Los Alamos National Laboratory was revealed by HP in the year 2005. In 2007, a sporadic occurrence of incidents causing system lock-up were due to existence of a flaw in the Translation Lookaside Buffer of multiple AMD Phenom processors series. The buggy TLB was disabled by initial BIOS and software workarounds resulting in an average 10%-15% performance degradations. The bug caused severe damage to the AMD reputation [265].

Finally, Reliability has evolved like design characteristic such as temperature dissipation, power consumption and performance for multi-core and many-core systems. The remaining portion of the article is organized as (3) Types of faults and metrics (4) Damage of physics (5) failures and fault-models of permanent faults (6) failures and fault-models of soft errors (7) fault tolerance methods in logic circuits for Digital CMOS ICs (8) fault tolerance methods for SRAM based FPGAs (9) fault tolerance methods at micro-architectural level (10) fault tolerance methods at architectural level (11) fault models and fault tolerant designs in TSV based 3D NOC many-core systems and (12) Core level fault tolerance methods for many-core systems. The scope and extent of the article is presented in section 2.

## 2 SCOPE AND EXTENT

We examine fault model, failure mechanism and fault mitigations techniques for soft error (single event upsets and transients), permanent faults and timing delay at combinational and sequential logic level (Digital CMOS and FPGA), micro-architectural and architectural (or multi-core) level. Fault mitigation approaches for SRAM based FPGA are examined only at logic layer and for the configuration memory. We examine only 3D NOC integration by through silicon vias, with relevant logical fault models and failure mechanism. At many-core level, we discuss only hardware-based approaches to detect and diagnose faults with in field reconfiguration capability thereby improving the reliability of 3D NOC many-core systems. We do not discuss device level enhancements, application specific approaches, routing approaches, algorithm level or program level mitigation approaches in this article. We also do not examine cache and main memory vulnerability in this survey.

## 3 TYPES OF FAULTS AND METRICS

In an electronic system, defects can be defined as the unbearable dissimilarity between the operational hardware and its conceptual design. Some defects are (a) package defects - seal leaks, contact degradation (b) Age defects - electro-migration, dielectric breakdown etc (c) Material defects - surface impurities, bulk defects (cracks, crystal imperfections) (d) process defects - oxide break-down, parasitic transistors, missing contact windows. At the abstracted function level, "defect" is represented by a Fault. Error is an incorrect outgoing signal produced by a defective circuit or system. Failure is a special case of error, in which the error deviates the system from the expected action. It is important to note that not all errors cause failures. In broad sense, the faults are classified as permanent faults and temporary faults. The temporary faults are further classified as intermittent fault and transient fault. We examine permanent, intermittent and transient faults below.

**Permanent faults: -** The defects in the silicon semiconductor material or metals used in the construction of a processor packages cause permanent faults. The permanent faults are irreversible and result in processor failure. The permanent fault rate or hard error rate and lifetime of a processor are inversely proportional to each other. Extrinsic failures and intrinsic failures are the two types of permanent faults or hard failures. Process and manufacturing defects are the primary reasons for the occurrence of extrinsic failures. The occurrence of extrinsic failure decreases over a period of time and an early detection of it is possible in the initial phase of processor lifetime. The surface roughness and contaminants on the crystalline silicon surface, for example, can cause dielectric breakdown [58]. Extrinsic failures are largely a derivative of the manufacturing process. The processors with extrinsic failures can be detected using approaches like Burn-in and voltage screening. The natural limitation of the material and intrinsic failures cause wear-out failures. Under a set of conditions, specified processor in operation would encounter intrinsic failures. The rate of intrinsic failures increases over a period of time. The best examples for intrinsic failures will be electromigration in interconnects, Time dependent dielectric breakdown (TDDB) in the gate oxides, and thermal cycling and cracking.

**Intermittent faults: -** The process variations, partial oxide degradation, in-progress wear-out and manufacturing residuals cause frequently Intermittent hardware faults that can occur aperiodically and non-uniformly for interval of time. In the lifetime of any processor, permanent faults prevail, at the same time environmental and process variations may activate or deactivate the occurrence of an intermittent fault.

**Transient faults or soft errors: -** The primary source for soft errors is extra-terrestrial (i.e., solar flares) and terrestrial (i.e., radioactive decay) occurrences. Terrestrial sources include the particles generated due to decay of radioactive impurities in the material used in packaging of the chip. In extra-terrestrial phenomena, the primary cosmic rays react with the earth's atmosphere via strong nuclear interactions, producing various particles which can induce soft errors. Other than three faults mentioned above, timing faults and design faults are discussed briefly here. Aging causes the transistor to slow down and would result in timing fault. Drastic variations in temperature and voltage would cause timing faults. The design faults creep in as the circuits are designed and synthesized, would cause impairment in functioning of the processor.

**Metrics: -** Failure rates can be given by Time to Failure (TTF). It is the first occurrence of fault or errors. Similarly, mean time Between Failures (MTBF) indicates the mean time that has passed between two faults or errors. Mean Time to Repair (MTTR) indicates the mean time required to repair an error after it is detected. Further, Mean Time to Failure (MTTF) is the time until the system encounters a failure once it is repaired. Failure rate are easily expressed using Failure in Time (FIT). One failure in a billion run-time hours is equal to one FIT. FIT rate of a system is the summation of individual FIT rates of all the components. For example, if a 6T-SRAM cell with the failure rate of 0.001 FIT/bit is used to design a 1 MB cache, then the total failure rate of the cache is 8389 FIT and the cache has an MTTF of about 4900 days. MTTF and FIT are inversely proportional as in Equation-(1).

$$FIT\ rate = \frac{10^9}{MTTF\ in\ years\ \times 24\ hours\ \times 365\ days} \quad (1)$$

114 FIT is comparable to 1000 years of MTTF. Chip designers have fixed FIT (or MTTF) target just like power budget. Further, handling both permanent and transient faults are important for reliability. Unlike soft errors, the permanent faults can be identified during validation and are fixed before the silicon chip is shipped. However, soft errors must be handled in the field. We discuss the sources and fault models for permanent faults and transient faults (or soft errors) in the 5 and 6 sections.

## 4 PHYSICS OF DAMAGES

As early as 1970, damages due to radiation in the aerospace applications and military environments were examined for the first time. Binder *et al.* [266] published technical report on soft errors, first of its kind in the year 1975. May *et al.* and Woods *et al.* of Intel presented a first paper at the International Reliability Physics Symposium (IRPS) in 1978 on "occurrence of soft errors at sea level" highlighting the soft error implication on DRAMs [267]. Rapid progress in CMOS has resulted in reduction of transistor size thereby radiation became the main cause of apprehension. The telluric applications encounter soft error effects induced by high energy particles such as thermal neutrons, alpha particles and high -energy neutrons [36]. Contemporary seminal research article throws light over evolving dangers of transients induced precisely by muons [169] and probable risk from electrons and gamma rays in the near future. The space applications encounter soft errors induced by high energy cosmic particles including protons and heavy nuclei. In this section, we examine three major damages (a) total ionization dose (TID), (b) displacement damage dose (DDD), (c) single event effects (SEE) on Digital CMOS and SRAM based FPGA.

**Total ionization dose: -** In the ionization process, the photons lose their energy and gets charged in the electronic devices when a beam of gamma ray photons encroach on the surface of the devices. The ionization dose is defined as the energy quantitatively needed to form electron-hole pair. The circuits at the physical level get exposed for a period of time to absorb the total energy available is called total ionization dose [12]. The gamma photons can precisely initiate the process of Ionization or can be triggered by secondary recoil particles emitted when photons are generated. Hence, charge escalation in the dielectrics of the devices affected by TID is observed. The silicon dioxide is the commonly used dielectrics in the transistors today. The charges are ambushed in the silicon dioxide when the high energy charged particles hit the silicon dielectrics of the transistor. This phenomenon can be described in two steps (a) electron-hole pairs are created by ionization of atoms by high-energy protons and electrons. Generated electron-hole pairs may create new electron-hole pairs if they have high energy possessed by them. Thousands of electron-hole pairs are generated by one focal proton or electron. The left-over afresh generated electron-hole pairs combine themselves again. The electron deftness is very much greater when compared to the deftness of holes in the silicon dioxide prevents electron-hole pairs combining themselves again. The applied potential across the dielectrics has quantitative implications on electron-hole pairs combining themselves again. The applied electric field will coerce holes and electrons to move in the differing directions. (b) Polaron hopping, a process where holes that do not combine themselves again will move through the dielectrics. The direction of the holes is pronounced by the potential applied across the dielectrics. The movement of holes in the silicon dioxide results in malformation of local electric field of the dielectric lattice. The traces of holes in the dielectrics at it moves and the charges of the holes combines with oxide strains generates polaron and the process is called polaron hopping [13]. The holes that are not ambushed deep in the oxide will have two options (a) leave the silicon dioxide, or (b) form an interface state near the surface of the silicon dioxide that would have substantial implication on the electrical properties of the devices.

**(a) TID in digital CMOS: -** Predictably, in MOSFET, gate oxides are the key victims of TID damages. Increase in the drain-to-source leakage in n-channel transistors due to uncontrolled or no control on charges movement, and is also the cause for decrease in current in p-channel transistors as a result of charges ambushing in the gate oxide that changes the threshold voltage of the MOSFET. Rapid CMOS improvements has resulted in transistor scaling the escalation of charges on the either side of the thin gate oxide layer is very small or absent [14]. Probably charges ambushed in the gate oxides will be compensated or wrecked by the electrons tunnelling through the dielectrics. Shallow trench isolation (STI) usage in the CMOS to avoid latch-up is the key giver for escalation of charges that would result in TID damage [15].

**(b) TID in SRAM based FPGA: -** The usage of STI and MOSFETS in large numbers in the CMOS portion of FPGA are vulnerable to TID damage. The level of damage caused by TID in SRAM based FPGA depends on (a) dose rate (b) type of radiation (c) the internal electric field including space charge effects [16] (d) device geometry [17,18], (e) operating temperature, time after irradiation (annealing or rebound), [19,20]. The charges ambushed in the silicon dioxide dielectrics results in ionization effects and it has major implication on significant parameters like (a) reduction in threshold voltage [14], (b) causes a deterioration of noise parameters, (c) reduction of drain-source breakdown voltage, (d) reduction in surface mobility, (e) decrease of trans-conductance, and (f) an increase of leakage current [21,22]. The radiation causes threshold voltage shift swiftly in N-type metal-oxide-semiconductor (NMOS) transistors when compared to p-type metal-oxide-semiconductor (PMOS) transistors. PMOS is inherently tolerant to radiation when compared to NMOS. In NMOS transistors, positive threshold voltage can either decrease or increase [23] and makes it more vulnerable than PMOS. The positive gate bias voltage forces the charge convoy to move towards the interface state near the dielectric; as the ambushed charge in the oxide ($Q_{ot}$) dominates when threshold voltage swings to the other side or reduces. As the charge near the oxide increases, threshold voltage can swing in the opposite direction. The shift in the threshold voltage in PMOS [24] is the outcome of holes existing as the charge carriers which are slower and less electrons are carried when compared to NMOS having electrons as carriers [25]. PMOS is two to three times less affected in terms of area when compared to NMOS for a given constant area of influence from radiation. Excess charges in gate oxide in NMOS will negatively shift the threshold voltage resulting in unbearable levels of drain-source leakage current draining. Contrary occurs in PMOS, reduction in the leakage current and rise in the threshold voltage is observed [26].

**Displacement damage dose** (DDD)**: -** The collisions of heavy energy particles like neutrons cause disturbance in the crystal lattice structure of atom. The focal particles hitting the material may loses its energy by means of non-ionizing process can cause atomic displacement and intrinsic damage or defect. Collision of neutrons can result in elastic scattering or inelastic scattering. The neutrons scatters in multiple directions as it collides with nucleus in elastic scattering. The nucleus hit by the neutrons gains momentum in term of energy and the neutron loses. As neutron collides with nucleus in inelastic scattering that results in forming compound nucleus and produces a gamma radiation in the de-excitation process. The capture effect is more expected to happen in low-energy neutrons and is more probable in high energy neutrons in elastic scattering. The ionization in the targeted material as result of neutron interaction will produce secondary particles. An alpha particle produced due to ionization of the target material will have high linear energy. DDD is generally caused by neutrons. The major and minor effects of particle interactions are summarized in **Table 2** [27-30]. The interstitial, divacancy, and vacancy are important classes of DDD [29]. Finally, DDD (a) modifies the electronic characteristic of semiconductor junctions, (b) forms irreversible damages and upsurges the number of recombination centres, (c) draining the marginal carriers, and (d) modifies the arrangement of the atoms in the crystal lattice of the material.

**Single Event Effects: -** The consequence of radiation, ionization process, and DDD on the electronic devices results in the generation of single event effects (SEE). The linear energy transfer (LET) is a happening known to take place when a pair of electrons-hole generates in the semiconductor device is proportional to the energy deposited in the material [31]. LET is measured in MeVmm$^{-1}$; energy subsumed material whose mass is proportional to the energy transfer is measured in MeVcm$^2$g$^{-1}$ [32]. The high energy particle passing through the semiconductor device without resulting in failure or not detected is the maximum LET known as Critical LET, or the LET threshold (LET$_{th}$). The outcome of LET generates electrons-hole pair manifests as a charge, is the lowest charge required to trigger SEEs is called as critical charge [33]. The hard errors and soft errors are two variants of SEEs [34] and SEEs classification tree is shown in **Figure 1.** The burnout effect from short circuit would be an appropriate analogy for the irreversible damages in the hardware structure caused by permanent fault or hard errors. A soft error is a flip in data bit or a temporary change in the signal that commonly occurs in the logic circuits. The partial functioning or faulty operating semiconductor device which continue to do so even when they encounter soft errors [35]. The notable design approaches exist which can detect and correct soft errors without additional power cycles in the electronic device.

**Soft errors: -** The electrons-hole pairs are created when the charged particles travel through a semiconductor material. The majority portion of the reverse bias PN junction is more vulnerable. The voltage/current transients are generated when the collected charge carriers drift towards the neighbouring node. The quick drift process results in the accumulation of charges followed by diffusion process [36]. The accumulation of charges drifted increases in the funnel-shaped extension of the depletion region; thus, charges collected at the node rises effectively causing a change in the turn-off state [37,38].

**Table 2 Major and minor effects of particle interaction**

| Particle(s) | Energy Band | Effects of particle interaction | Major effects | Minor effects |
|---|---|---|---|---|
| **Alpha particles** | Typical 4- 8 MeV | Coulomb attraction | Ionization phenomena | |
| **Photons** | <0.1 MeV | Photoelectric effect | Ionizing phenomena | Displacement damage |
| | 0.3 - 3 MeV | Compton effect | | |
| | > 1.024 MeV | Pair production | | |
| **Neutrons** | ~0.025eV | Slow diffusion and capture by nuclei | Displacement damage | Ionizing phenomena |
| | < 10 MeV | Elastic scattering | | |
| | >10 MeV | Elastic, inelastic scattering, and secondary charged reaction products | | |

**(a) Single Event Transient** (SET): - A small glitch in the logic circuit or changes in the state of the flip-flop or memory elements are the symptoms of SET occurrence [39]. SET are not harmful because they can be detected and corrected. The high-speed clocks enhance the probability of capturing the transient pulses [40]. The static riming analysis would not be helpful in seizing SET which is not a synchronous phenomenon.

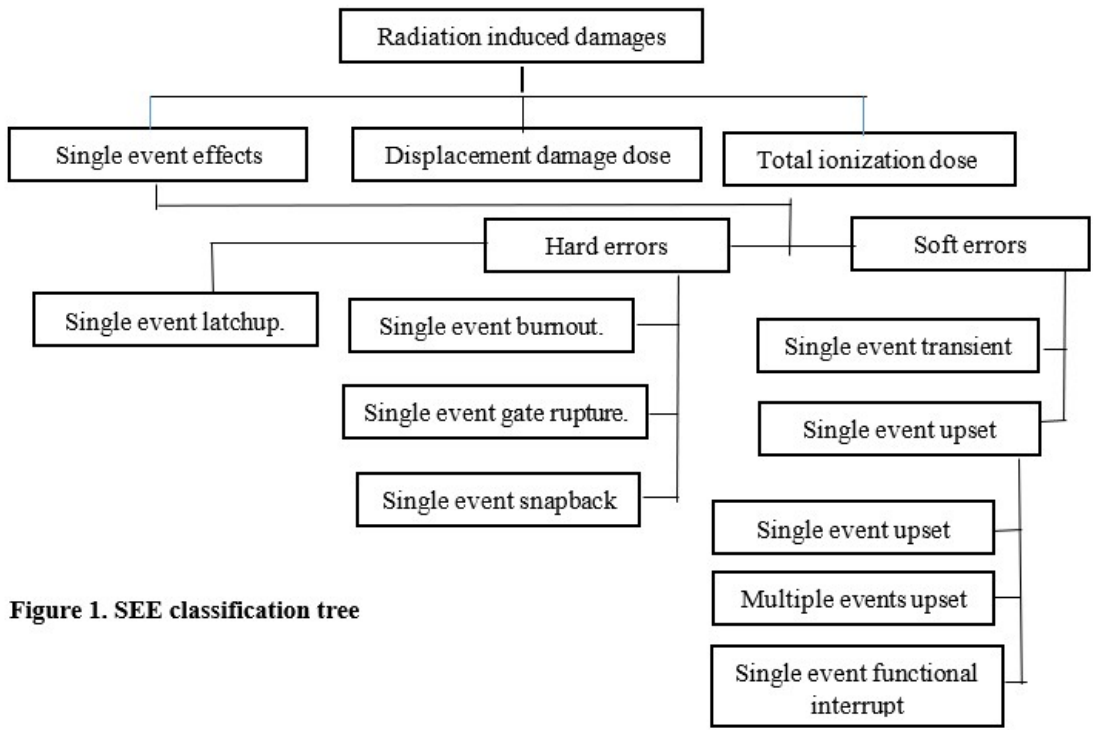

Figure 1. SEE classification tree

**(b) Single Event Upset** (SEU): - SEU has a negatively influencing ability on CMOS, memories, FPGA, and bipolar junction transistor. The high energy particles i.e., ions, protons, recoil particles of neutrons nuclear reactions, with ample energy causes SEU that modifies the state of the semiconductor device thereby results in an inducement of an error. The delicate parts of an electronic component are ionized by charged particles thereby generates electrons-hole pairs [41]. The semiconductor materials are heavily ionized by high LET particles as they transfer energy while traversing the miniscule path. SEU affects CMOS devices. In MOSFET transistor, an energy ambushed at the delicate node of the device results in the increase of drain current is observed. It is found that current induced is in micro-amperes and would last for $10^{-9}$ seconds. The induced current pulse is in microamperes and its last for nanoseconds. The critical charge is estimated using the Equation-(2) mentioned below. The $LET_{th}$, is a phenomenon that depends on IC fabrication technology.

$$Q_{crit} = \int_0^{tf} I_d(t)dt \qquad (2)$$

For 0.5 μm node size, to produce a SEU, the critical charge in the range of femto coulombs is sufficient. The sensitivity of SEU is estimated using the cross-section of the exposed material and is measured in $cm^2$/bits or $cm^2$/device.

**(c) Multi-Bit Upset** (MBU): - A particle one or more hit on a memory cell can splash or knock over the neighbouring cells that would result in multiple bit flips or upsets. Three important reasons that forms the basis for multi-bit upsets are (a) The spallation reactions caused by key particle in the chip results in products of varying energy level that cause upsets in memory cells (b) one or more memory cells will be under the influence of the charge that is ambushed in the diameter of the cylinder resulting many SEUs occurring (c) angle of particle creating an impact in one or more cells will result in multi-bit upsets [42].

**(d) Single Event Functional Interrupt** (SEFI): - SEFI is a variant of SEU. SEFI occurs in the control logic circuits. The device loses its operational characteristics when SEFI strikes the control units. It terminates the functioning of the device and the standard operation is resumed by an external signal from the peripheral circuits [43,44]. The specific circuits that handle power-on-reset, configuration memory, joint test action group (JTAG) or select-map communications port are the primary targets of SEFI in SRAM based FPGAs [45,46].

**Hard Errors: -**

**(a) Single Event Latch-up** (SEL): - Digital CMOS ICs are the prospective devices that get shattered by SEL [47]. The scrounging thyristor pnpn (or npnp) can be conceptualized by two parasitic PnP and NpN bipolar junction transistors. The usual progression in operations of a thyristor will be affected by a reverse-bias at well-substrate junction. The functioning of the thyristor is stopped. Thus, between thyristor anode and cathode the flow of current is blocked. The compliance of the following condition enable thyristor triggering when a high LET particle strikes. 1. The opportunistic (or parasitic) transistors with gain $β_{npn}$, $β_{pnp}$ product should be higher than 1. 2. Injection of higher dosage of current at base-emitter junctions of the opportunistic (or parasitic) transistors. The sustenance in the current levels is ensured by power supply thereby continuing the latch process. The high flow of current in the low impedance path causes excessive heat dissipation and destruction when latch-up is detected in an activated thyristor. Thus, resuming normal operation is attainable (a) when power supply is terminated (b) deactivating parasitic thyristor.

**(b) Single Event Burnout** (SEB): - VDMOS (or DMOS) is a high-power MOSFET transistor affected by SEB. The high-energy particle takes a course of path that traverses through bipolar junctions of the transistors. The high density of electrons-hole pairs is generated when particle hits take place. The high drain-source voltage is juxtaposed with current density of high order $10^4$ $A.cm^2$ gets generated after high energy particle strikes [48, 49]. In voltage drops at the base-emitter junction in the opportunistic (or parasitic) transistors that is being turn on thereby resulting in collection of charges at the collector. At junctions in transistor, excessive heating is the outcome of collector current results in device burnout.

**(c) Single Event Gate Rupture** (SEGR): - In power MOSFET transistors, breakdown of the dielectrics by high energy particles is a matter of major concern [51]. The electrons-hole pairs are generated at the gate oxide as the particle traverse through the dielectrics and agility of the electrons is under the clout of the electric field that exist between the drain and gate. The electric field at the dielectrics increases as the positive charges accumulates at the silicon dioxide interface. It results in the rise of the leakage current due to collector current and field density. The temporary (or transient) commotion is sufficient to release the ambushed charges at the dielectrics. It causes an increase in the temperature in the neighbouring regions of the silicon dioxide interface. Rise in the temperature ruins the gate oxide [50,52]. The non-volatile memories like EEPROMS are affected by SEGR.

**(d) Single Event Snapback** (SES): - The Digital CMOS and Silicon on insulator (SOI) based devices are the major victims of SES [53,54]. NMOS transistors are largely affected by SES. The high LET particles turn on the parasitic bipolar junctions by ionizing heavily on MOSFET structure. Disastrous effects of SEL and SES are similar. In SES, normal operations can be resumed by external reset signals without decreasing the power supply, unlikely in SEL. If the local current densities are high, then SES will be more devastating. CMOS circuits are vulnerable to radiation. Both TID and DDD damage the devices with bipolar and MOSFET based transistors. Shrinking dimensions of the MOS transistor make devices more vulnerable to SEU. The operation frequency of the devices and effects of SET are proportional to each other. The thin gate oxide in MOSFET transistors of the submicron CMOS devices have become more resistant to ionizing radiation [55]. Further reduction of the node size will be source of complications for transistors consisting of thin gate oxide. The dielectrics or silicon dioxide interface with higher electric field will be inflicted by the SEGR [56][57]. The summary of SEE on different components with respect to technology are given in **Table 3**.

**Table 3 Implications of SEE on Components with different technology**

| Technology | Family (digital / analog) | Function | Single event effects |
|---|---|---|---|
| CMOS, BiCMOS and Silicon on Insulator | Digital | SRAM based FPGA | SEL, SES, SEU, MBU, SEFI, SET |
| | | Flash /EEPROM | SEL, SES, SEU, SEFI |
| | | Microprocessor/ Microcontroller | SEL, SES, SEU, MBU, SEFI, SET |
| Power MOS transistor (like VDMOS or DMOS) | | | SEGR, SEB |

## 5 FAILURES AND FAULT MODELS OF PERMANENT FAULTS

In CMOS devices, intrinsic failures and extrinsic failures give rise to permanent faults. The intrinsic failures or wear-out in the CMOS devices determine the lifetime reliability or life-span of a processor. The silicon dioxide used as a dielectric in the CMOS device manufacturing weaken over a period of time results in a intrinsic failure. The book by Segura *et al.* and Hawkins *et al.* [89] discusses fault models in greater detail. In this section, fault models and intrinsic failure mechanism for hard errors or permanent faults encountered by processors are only considered for discussion. The failure mechanism can be broadly classified as (a) oxide failure modes, and (b) Metal failure modes.

**Oxide failure modes: -**

**(a) Negative Bias Temperature Instability: -** In the digital integrated circuits and ultra deep-sub-micron clones, Negative bias temperature instability (NBTI) is considered to be a significant reliability challenge [66,67]. The pMOSFET is switched on during high temperature (between 100 °C and 150 °C) results in a happening known as NBTI. The defects are inflicted on the semiconductor device consequent to irreversible diminishing current drive and shifts the threshold voltage ($V_{th}$) [67, 68]. NTBI is modelled and described using hydrogen release model. Under conditions like high temperature and application of voltage, the high energy electrons-hole pair strike the silicon dioxide interface. It results in the disintegration of silicon-hydrogen bonds thereby discharge of hydrogen atoms takes place, also a by-product of electro-chemical reactions at the gate dielectric interface. The free hydrogen atoms join with nitrogen or oxygen atoms that generate positively charged ambushes or pits at the dielectric interfaces. Above phenomenon makes pMOSFET transistor threshold voltage to shift non-positively, an after-effect of diminishing mobility of holes. The term "Instability" in NBTI denotes to the changes in the threshold voltage over a period of time.

**(b) Hot carrier injection** (HCI): - Very high electric field at the drain of the MOSFET transistor expedites the occurrence of HCI. The high frequency performance and timing logic in the circuits undergo transformation due to HCI. Hot carriers scorching near the very high electric field next to drain is detrimental to the MOSFET transistors that results in ultimate deterioration in the parameters which characterises the device. Theses carriers are known as Hot carriers because they possess high energy. The $I_{sub}$ denotes the substrate current, will rise due to admittance of Hot carriers in the substrate area. The carriers with sufficient energy levels (i.e., 4.6 eV or higher for holes **or** 3.1 eV or higher for electrons) can definitely pass through the gate-oxide barrier and thereupon induce defects [65]. The estimated drain saturation current ($I_{Dsat}$) is used directly in estimating the deterioration due to HCI, is a standard practice followed for many years. Because $I_{Dsat}$ is very important parameter for MOSFET transistor. HCI induced defects and its implication on the performance of the circuits during the normal operation (MOSFET in saturation mode) can be adjudicated using drain saturation current. In the Equation-(3), drain current is denoted by $I_D$; transistor width is denoted by W; substrate current is denoted by $I_{sub}$; transistor lifetime degradation is denoted by τ.

$$\tau = C \, \frac{\frac{W}{I_D}}{\frac{I_{sub}}{I_D}} \qquad (3)$$

**(c) Time-dependent dielectric breakdown** (TDDB): - The thin dielectrics in the MOSFETs is made out of silicon

**8**

dioxide that exist in non-crystalline and amorphous state. The source and the drain of the MOSFETs are made crystalline silicon doped with contrasting polarity with respect to the substrate. The control terminal of the MOSFET is the gate oxide. The drain collects the electrons or holes generated at the source. The gate oxide secludes the movement of the electrons-hole pair that results in the collection of charges at the silicon dioxide interface when electric field is decreased (increased) in PMOS(NMOS) transistors. Between the substrate and the gate oxide terminal, a conducting path is formed. The electric field at the gate oxide will not be able to regulate charge flow between source and drain. Unregulated current flow would make the device inoperative [58,60].

**Model: -** The voltage applied, electric field and the temperature determines the infallibility of the dielectrics in the MOSFET. TDDB is closely related to electric field, its inverse and voltage at the gate. Wu *et al.* [64] proposed a model that estimates the longevity with respect to TDDB for ultra-thin gate oxides which relies on voltage and the temperature that causes exponential deterioration. In the Equation-(4) [64], b, a and $T_{BD0}$ are analytically estimated constants.

$$MTTF_{TDDB} = T_{BD0}\ e^{(\frac{a}{T} + \frac{b}{T^2})} \qquad (4)$$

**Metal failure modes: -**

**(a) Stress migration: -** The conductor metal atoms in the interconnect wires relocate due to mechanical stress and this happening is known as stress migration. The disturbance in the crystal lattice of the semiconductor material causes intrinsic stress that results in stress migration. The varying thermal stretches of incomparable materials in the device causes thermo-mechanical stresses that results in stress migration [60,62]. In TSV, stress migrations results deformation in the molecular structure of the metal wires or creates holes (or pits) in the metal wires, forming a void by sneaking and meeting in a single location [150].

**Model: -** The thermo-mechanical stresses form the basis for the failure model on stress migrations. The varying thermal stretches in incomparable materials in the device is caused by stresses. The σ denotes mechanical stress due to varying expansion rates, and is directly related to the variations in temperature. The variations in the temperature are estimated with regard to metals of stress-free temperature. As the thermal stress reduces to the negligible, metal deposition temperature is the stress-free temperature of the device when the metal is encroached on it. The thermo-mechanical stress exists at all temperatures other than the metal deposition temperature. The MTTF for the stress migration $MTTF_{SM}$ is estimated [60] using the Equation-(5) as mentioned below. In Equation-(5), $A_{SM}$ and n are constants; value of n ranges in between 2 and 3 [60,62]; activation energy for stress migration is denoted by $E_a$.

$$MTTF_{SM} = A_{SM}\sigma^{-n}e^{\frac{E_a E}{kT}} \qquad (5)$$

**(b) Electromigration: -** The surge in the temperature and flow of electrons would advance electromigration in the metal wires or interconnects. It creates fissures or hollows in the metal wires or interconnects. The plenitude of conductor atoms in the copper and aluminium metals transmitted over interconnects is due to the propulsion in the flow of electrons. The conductor electrons contribute their acquired energy to the metal's atoms of the interconnect wires; thereby an electron convoy or storm is formed, which forms a aspiring force for the bottom-line metal atoms to flow as guided by the electron flow. Relocation of metal atoms as it happens results in the metal atoms diminishing in one section and heaps up in the other section. It further complicates by coalescing and expanding fissures or hollows to many regions of deficit metal atoms resulting in (a) rise in the resistance of the wires, (b) open circuits, and others challenges. In the areas where conductor atoms chunks are found, formation of extrusion would produce shorts between side-by-side interconnect wires produces failures in the circuit.

**Model: -** The Black's original electromigration Equation-(6) [59, 60, 61] is the presently established model to estimate the mean time to failure (MTTF) for electromigration ($MTTF_{EM}$). In the Equation - (6), Boltzmann's constant is denoted by k; activation energy for electromigration is denoted by $E_a$; current density in the interconnect is denoted by J; critical current density required for electromigration is denoted by $J_{crit}$; absolute temperature in Kelvin is denoted by T; constants $A_{EM}$ and n; value of n relies on the material of an interconnect may range between 1 to 2 [60, 61, 62];

$$MTTF_{EM} = A_{EM}\ (J - J_{crit})^{-n}\ e^{\frac{E_a}{kT}} \qquad (6)$$

In the Equation – (7), *f* denotes the frequency of clock; current density denoted by J of a metal line can be correlated to the switching probability (denoted by p) of the line [63]; capacitance denoted by C; width denoted by W; thickness denoted by H;

$$J = \frac{CV_{dd}}{WH}\ x\ f\ x\ p \qquad (7)$$

**Fault model (logical level) for defects:** - The bridging defect (or fault) model, cross-point defect (or fault) model, stuck-short or stuck-open defect (or fault) model, stuck-at defect (or fault) model are the different types of fault models that are commonly used to model the defects in the device. The allocation of constant values (0 or 1) to the signal lines in the circuit forms the basis for modelling the defects in the Stuck-at defect (or fault) model. The latches and flip-flops have input and output lines, are the signals lines stuck by defects apportioned with constant values. The stuck-at-1 and stuck-at-0 faults are also called as single stuck-at faults are the commonly used fault models for defects.

## 6 FAILURES AND FAULT MODELS OF SOFT ERRORS

The transient faults infliction in CMOS devices is caused by radiation that comes from two cradles. They are neutrons from atmosphere and materials used for packaging generate or release alpha particles. The solar particles and galactic particles are the primary constituents of the primary cosmic rays. The proportion of heavier atomic nuclei, alpha particles and protons present in the galactic particles are 2%, 6% and 92% respectively. The neutrons in the earth's atmosphere are predominantly the neutrons that are the constituent particles of galactic particles. The soft error rate (SER) in the CMOS devices is governed by neutron's energy and flux. The secondary cosmic rays are the outcome of collision course of primary cosmic rays with the earth's atmosphere. The neutrons, muons and pions are the secondary particles. The debilitation of muons and pions happens in $10^{-3}$ and $10^{-9}$ seconds as they lose their energy. The cosmic rays that are terrene ultimately strike the surface of the earth with their constituent particles. The CMOS device are affected by earth-bound neutrons with 10Mev or above energy levels. The contaminants with radioactive property found in the packaging material produce alpha particles. The radiation induced transient faults are discussed in greater detail by Shubu Mukherjee book [65]. In the following subsections, interaction of neutrons and alpha particles with silicon dioxide and boron are briefly presented.

**Neutrons:** - The interaction of low energy cosmic neutrons with boron nuclei is the seedbed for ionizing particles in semiconductor devices. P-type dopant such as boron is used widely. The neutrons are involved in inelastic collisions, first silicon recoil (or Li recoil in the case of interaction with boron nuclei) and secondary particles are generated which finally result into generation of electron-hole pairs as Impact of a higher energy neutron results into higher energy recoils. Each neutron can generate about 10x more electron-hole pairs compared to an alpha particle [69]. The charge density per distance travelled for silicon recoils (25-150 femto Coulomb/µm) is significantly higher than that for alpha particles (16 femto Coulomb /µm) and hence, neutron strikes have higher potential to upset a circuit [72]. Typically, a neutron with 200 MeV energy, generates a recoil that has stopping power of 1.25 MeV/µm and maximum penetration range of 3 µm [72]. One such particle strike can deposit total charge of 55.7 femto Coulomb [73].

**Alpha particles:** - The two neutrons and two protons join together to form an alpha particle. Alpha particles come from residual radioactive impurities (e.g., Uranium (U238), Thorium (Th232), and Lead (Pb210)) in the packaging material of a chip [69, 70, 71]. Packages, which use solder balls for the power supply and I/Os, are particularly vulnerable to soft errors. In order to reduce the alpha induced soft errors highly refined materials can be employed for packaging materials. Alpha emitting materials have an emission rate of 0.0003-0.0017 alphas/cm2 - hr [69, 71]. In an inelastic collision involving an alpha particle, electron-hole pairs are generated through direct ionization in silicon and 10 MeV of energy possessed by an alpha particle has a stopping power (LET) of 100 Kev/µm and can generate approximately 4.5 femto Coulomb /µm of charge [69,72].

**Implications of soft errors:** - It was found in the early 2000, Ultra SPARC II workstation break down at frightening rate. The origin of this challenging issue was tracked, and was found that SRAMs purchased from IBM encountered SEUs. As a result, Sun had to switch memory vendors and also designed error detection and correction mechanisms for their caches [74]. Due to high solar activity in 2003, 28 satellites were damaged, out of which 2 were unrecoverable [75]. The Qantas Airways operated an Airbus A330-303 destined towards Singapore from Perth experienced a setback operationally. The aircraft without any noticeable and uncontrolled behaviour inclined towards one side resulting in harming nine crew members and 110 passengers [76]. In the decades to after 2020, it is expected to have 50 billion networked devices [77] and number of chips around us are increasing due to explosion of semiconductor devices usage in everyday life. Per user, rise in the transistor usage is inferred as per user rise in the soft errors in the near foreseeable future.

**Soft errors modelling:** -

**Soft errors modelling at circuit level:** - The soft error rate is estimated for processors in two steps (a) Intrinsic Failure in Time (or device level soft error rate) is estimated for the devices and circuits (b) de-rate the intrinsic FIT rate using vulnerability factors of the devices and circuits. Two-steps are involved in estimating intrinsic FIT rate of an element in the circuit and they are (a) neutron or alpha particles strike generate charges known as critical charge Qcrit which can demolish the oxide barrier that cause breakdown in the circuit, and (b) estimated Qcrit is aligned or approximated with comparable FIT rate of an element in the circuit. The vulnerability factor of the circuit element is used to de-rate the intrinsic FIT rate estimated.

**Aligning Qcrit to FIT: -** Estimated Qcrit for an element in the circuit is aligned or calibrated to a soft error rate measured in FIT. Three well-known models that level Qcrit to soft error rate are presented below.

**(a) Neutron Cross-Section** (NCS) **Method: -** Taber *et al.* and Normand *et al.* [80] proposed NCS method that directly relates the parameters like energy and flux of the neutron with upset rate of the device. This method avoids using SV parameter as well as Qcrit. The Equation-(8) states upset rate estimation using NCS method. In this method, neutron cross section relies on the probability of upset produced by neutrons with energy (denoted as E-neutron) while interacting with other materials. For NCS method, device specific accelerated neutron test data is required to estimate probabilities. A comprehensive experimental test is needed to estimate probabilities and Qcrit, for their effective use in NCS method.

$$Upset\ rate = \int (\sigma \frac{dN}{dE})dE \quad (8)$$

**(b) Burst Generation Rate** (BGR) **Method: -** Ziegler *et al* and Lanford *et al.* [79] proposed an approach named BGR method grounded on two metrics (a) neutron-induced recoil energy (E-recoil) and (b) sensitive volume (SV). According to BGR method, an upset is a happening when a charge is created by neutron-silicon interaction inside the sensitive volume that results in a burst or shatter the device and quantitively larger than Qcrit. In the Equation-(9), E-recoil estimation is mentioned. The upset rate estimation as per BGR method is stated in the Equation-(10). In the Equation-(10), collection efficiency is denoted by Qcoll; neutron energy is denoted by E-neutron; differential neutron flux is dN/dE; BGR is a function of neutron energy and recoil energy. The experimental data is used to perform necessary summation of BGR function. The BGR experimental data is computed using heavy ion testing. The challenge in BGR method is to estimate SV that needs well-founded information about the chip layout.

$$E-recoil = Q_{Crit}\ x\ 22.5 \quad (9)$$

$$upset\ rate = Q_{coll}\ x\ SV\ x\ \int_{E-neutron}(BGR(E-neutron, E-recoil)\frac{dN}{dE})\ dE \quad (10)$$

**(c) Hazucha and Svensson Model: -** Hazucha *et al.* and Svensson *et al.* [78] proposed an equation as that models the Qcrit mapped to soft error rate of the circuit using flux of the neutron, area, and Qcoll (charge collection). The model's Equation-(11) is state below. In the Equation-(11), collection efficiency is denoted by Qcoll; diffusion area is denoted by area; neutron flux is denoted by flux exposed onto to the circuit. Constants and Qcoll are analytical constants estimated using accelerated tests. Precise process generation and new IC fabrication technology requires a necessary additional requirement to effectively compute circuit soft error rate using Equation-(11).

$$Circuit\ SER = Constant\ x\ flux\ x\ area\ x\ e^{-\frac{Q_{crit}}{Q_{coll}}} \quad (11)$$

**Aligning Qcrit to FIT using simulation models: -** Murley *et al* and Srinivasan *et al.* [81] proposed simulation-based modelling for the neutron strikes that result in charge collection in the semiconductor devices. The circuit is supposed to breakdown, when simulation results indicate charge collection larger than Qcrit. Using the known values of flux of the neutron, likelihood of the upsets can be determined using simulation and can be quickly changed to FIT rate. The comprehensive information about the physics of neutron interaction and node technology is required for creating simulation model. At last, computed intrinsic FIT rate of the circuit has to be de-rated using an applicable vulnerability factor to estimate circuit level SER. One such vulnerability factor is discussed below.

**Timing vulnerability factor** (TVF)**: -** The TVF [82] is a time fragment in a circuit susceptible to SEUs. The TVF relies on setup-time of latch or flip-flop. TVF of a SRAM cell is 100% as a result of any SEU that can modify the value stored in the SRAM cell during a clock cycle. Nonetheless, TVF for synchronous(asynchronous) clocked elements like flipflop (latches) can be lesser than 100%. To estimate the circuit-level soft error rate, it uses already estimated device level raw SER of the circuit de-rated by applicable vulnerability factors. One such factor is TVF, is multiplied with intrinsic FIT rate to estimate the circuit soft error rate. TVF relies on the propagation delay in the forward logic path of the circuit.

**Modelling soft error at architectural level: -** At the architectural level, soft error is modelled using Architectural Vulnerability Factor (AVF) [83]. The user visible errors of programs creep up due to fraction of faults quantified as AVF. If the AVF of a bit is high, then it that means that higher the susceptibility of the bit to cause errors. The fraction of bit flips that cause user visible errors in program outcome is given by bit's AVF. AVF for the Program counter is 100% and for branch predictor is 0%. Architectural Correct Execution bits or ACE bit are those that are directly visible to a programmer. Any change in this bit would affect the output of the program and they affect the correct path instruction execution. Predictor structures, Miss peculated state and invalid state are Micro architectural Un-ACE bits. NOP instructions, prefetch instructions, predicated false instructions and dynamically dead instructions are architectural Un-ACE bits.

In the Equation-(12), the AVF for hardware structure with N bits is mentioned below.

$$AVF_{structure} = \frac{\sum_0^n \frac{ACE\ Cycles\ for\ bit\ i}{Total\ cycles}}{N}$$
$$= \frac{mean\ number\ of\ ACE\ bits\ in\ structure\ in\ a\ cycle}{Total\ number\ of\ bits\ in\ a\ structure} \quad (12)$$

In the Equation-(13), the AVF of a processor is related with the FIT rate as mentioned below.

FIT = FIT$_{Raw}$ X TV F X Size (structure) X AV F     (13)

It obvious from equation above, FIT is a de-rated value by vulnerability factors and it provides a conservative estimate of processor's reliability due to AVF component.

**Silent data corruption (SDC) and detected unrecoverable errors (DUE):** - At bit level, SDC and DUE are caused by soft errors. Polluted data bit that goes unnoticed by the user are harmless and is not included in the computation of SDC. However, polluted or tainted data bit that ultimately produces a visible error that the attention on the SDC gains momentum from the user. Occurrence of DUE results in the system crash and the soft error is detected thereby prevents the corruption of the data. DUE is less harmful when compared to SDC. For example, SRAM storage cell has a SDC FIT rate stated as SDC FIT = Intrinsic FIT x TVF x SDC AVF, Intrinsic FIT rate of a cell denotes device level soft error rate, normal AVF is denoted by SDC AVF. Identically, DUE FIT = intrinsic FIT x TVF x DUE AVF. From the above de-rated products of failure rate, if a bit is protected with parity, then SDC AVF =0. If the bit is protected from error correcting, then DUE AVF=0. In turn DUE FIT is nearly zero. Logical deduction is that the error detection forms the basis for relative quantifying DUE AVF but the same is not applicable for SDC AVF.

Chip-level DUE FIT and SDC are estimated using Equation-(14) and Equation-(15) respectively.

$$Chip\ level\ DUE\ FIT = \sum_{i\ over\ all\ storage\ cells} DUE\ FIT\ of\ cell - i \quad (14)$$

$$Chip\ level\ SDC\ FIT = \sum_{i\ over\ all\ storage\ cells} SDC\ FIT\ of\ cell - i \quad (15)$$

**Impact of New Technologies: -**

**Silicon on Insulator (SOI):** - In the SOI devices, thinner silicon interface layer regulates the charges collected when hit by a neutrons or alpha particles, unlike the case in CMOS devices. Experiments on partially depleted SOI SRAM devices reported 5x reduction in soft error rate [84, 85]. A fully depleted SOI can further reduce the soft error rate by almost eliminating the silicon layer. However, this improvement in sequential and combinational logic is unclear.

**FinFETs and Multigate-FET Devices:** - In FinFET devices, charges do not accumulate near the drain region because the channel is the place of conduction and the majority of the charges disintegrates in the substrate region. It is worth observing that for the same technology node, the Qcrit of FinFET SRAM and planner CMOS SRAM are same. However, due to reduced charge collection, compared to planned CMOS device, 15x reduction in soft error rate of Tri-gate FinFET devices has been reported using device simulations at terrestrial flux [86]. Proton beam testing of the 22 nm Tri-Gate SRAM and sequential logic devices observed 1.5 – 4x reduction in soft error rate compared to 32 nm planner bulk CMOS [87]. Contrary, IMEC manufactured Tri-gate devices when tested using laser and heavy-ions, it was found that the region susceptible to charge collection in FinFET device is substantially larger than the existing structure, thereby rises the likelihood of SEUs occurrence in the cell [88].

**Parameters influencing the SER:** - We provide comprehensive summary of parameters that affect the soft error rate in the **Table 4**. The altitude and geographical location determine energetic particle flux. Not all particles cause soft errors, to cause soft error impacting particle must carry enough energy and it has to transfer its energy to generate enough charge to cause a fault, this energy transfer depends on the particle incident angle and its charge production capability. Because of these factors neutrons with less than 10 MeV energy are harmless. The location of particle strike determines how much charge will be deposited. The doping concentration along with the track length and track angles of the particle also affect the charge collection capacity. Nodal charge determines the Qcrit which is exponentially related to the soft error rate. Apart from that in circuit or microarchitecture domain the operating voltage, frequency, temperature and parametric variations also affect the soft error rate.

**Table 4 Parameters affecting the Soft Error Rate**

| Level of study | Parameters |
|---|---|
| **Striking particle** | Particle energy, Particle flux, Incident angle and energy, charge production capability |
| **Device or Material** | Position of impact, Track lengths, Track angles, stopping power or LET, Doping Concentration, |
| **Circuit (Combinational and Sequential)** | Nodal Capacitance, Sensitive area, Critical charge (Qcrit), Operating voltage, Frequency, Temperature, Parametric variations, Masking rate (Electrical, Logical and Timing masking) |
| **Microarchitecture** | Micro architectural masking rate (e.g., Dead instructions) |
| **Chip** | Packaging material, Process technology |
| **Environmental** | Altitude, Geographical location |

## 7 FAULT TOLERANCE METHODS IN LOGIC CIRCUITS FOR DIGITAL CMOS IC

Earlier, circuit engineers gave more emphasis for SEU mitigation techniques to protect the memories and ignored the logic paths comprising latches/flip-flops and gates (AND, OR and NAND) [117]. The soft errors impact on the logic datapaths were more explicit as the supply voltage and technology node were scaled down at 65nm node and further downsizing [90, 91]. The sequential circuits are vulnerable to SEU, SET and timing faults. The robustness of the sequential cell depends on spatial, temporal redundancy or combined method. The sturdiness of the sequential elements is presented in the next sub section. The combinational circuits comprising gates are less susceptible to soft errors. Masking effects suppress radiation induced faults so that not every generated propagate and produce errors. The logical masking, temporal derating, electrical masking and functional masking suppress the SEU induced by radiation and do not allow SEU to propagate and produce errors. The following section describe the masking mechanism.

**Functional Masking: -** Single event transient or single event upsets may modify the state order of the circuit. But the application function may have only negligible effect. For example, Single event upset may cause single pixel change in video stream is highly insignificant. This derating is called functional derating. Functional derating requires understanding of the application and impact of Single event upset can be classified as significant or insignificant with respect requirements.

**Logical Masking: -** A fault generated, if it cannot propagate through gates in the combinational circuit then it is logically masked or suppressed. The single event upsets can be masked in the cycle as it occurs or logically masked several cycles later.

**Temporal Masking: -** In sequential circuits, in order to propagate a fault to downstream flip-flop, they have to be sampled by sequential latch or flip-flop. For a single event upsets to be sampled in a flip-flop, it must sustain its presence in the clock period to meet the set-up time of one or more flip-flops. But the SEU will be masked. It is clear that available slack time increases so as the masking of single event upsets will probably increase. Single event transient is masked temporally and are called as latch window masking.

**Electrical Masking: -** Weak Single event transient on the device, does not cause the voltage to cross the threshold to create a fault and is masked. Due to their limited slew rate and intrinsic capacitance the logic gates behave as low pass filter and prevents very thin pulses. This capability is called electrical de-rating.

**Fault mitigation techniques: -**

**Triple Modular Redundancy: -** It is a standard practice in the majority of Triple Modular Redundancy (TMR) designs at the logic circuit level to triplicate the flip-flops. TMR gives ammunition against single event upsets and is habitually used in designing fault tolerant FPGAs. In TMR, absence of the redundant clock remains a concern, transients striking the clock circuits will disturb or make all flip-flops malfunction. Redundant clock circuits are incorporated in the hardware designs meant for space applications. In the triplicated clocks with the phase difference T for a period of time, transients with smaller duration when compared to phase difference is masked thereby preventing further propagation to downstream flip-flops. At last, transient with pulse 2 times larger than phase difference will have implications over timing capability of the clocking circuit.

**Parity codes: -** Data stored in the DRAM cells are safeguarded by using parity codes. Parity codes are also used to shield flip-flops thereby resistant to single event upsets is confirmed. The tremendous setbacks to parity codes usage is the extra-computational time to find the parity influences the critical path badly. Additional computation penalty incurred in the parity generation which is one half of an XOR gate per bit. Parity checking on the receiving side, incurs cost penalty of one half of a XOR gate in addition to comparison operation performed with the stored parity. The additional flip-flop that stores parity implies an additional cost is compensated by safeguarding all other flip-flops. To gather collective information about one or more errors emanating from one or more clusters of parity bits, an OR tree is needed with known cost penalty. Parity codes mask single error upsets in the memory elements. Established positioning methods prevents flip-flops from multiple bit upsets in memory elements or flip-flops.

**Fault tolerant cell designs: -**

Some of earliest radiation hardened cells are Heavy Ion Tolerant (HIT) cell [92] proposed by Bessot *et al.* and **Dual interlocked storage cell** (DICE) [93] proposed by Calin *et al*. The DICE has become standard for hardened flip-flops, and its layout were optimized and they were applied in industry to provide power sustained sturdiness by LEAP-DICE [94]. The formation of a feedback loop between two redundant latches joined back-to-back forms the basis for redundancy-based cell hardening design in DICE cells. When particle strike results in upset in any storage nodes, it will propagate in one direction only. The feed from the other direction restores the state of the upset node. DICE provide protection against the SEU but not for SET and timing faults. Simultaneous strike on two sensitive nodes will result in the correct value stored [95]. Seifert *et al.* [96] presented experimental results that indicate upsurge in the separation of sensitive nodes results in drastic fall or exponential fall in the soft error rate. Hazucha *et al.* [97] and Berg *et al.* [98] presented the underlining limitations in the design of the DICE cell. A transient disturbance can be observed at the output in between a DICE cell faces a single event upset strike and the recovery of the cell state takes sufficient time. In the master-slave mode operations of a DICE cell, invulnerable transistors in the kth slave stage affected by particles which possess ionizing capability, may undergo state change or upsets and even if it could be recovered then a transient is likely to be observed in the downstream flip-flops and at the output Q. The implications of singe even transients in the latch is gaining more importance [99] under higher operating frequency conditions thereby the need for Temporal De-Rating (TDR) has diminished over a period of time. The pass transistors in the DICE cell drives the clock are more vulnerable thereby single event upsets may change the state of the bits stored or data stored is polluted. The polluted pass transistor can cause (a) glassiness or vulnerability in the cell, obviously store polluted values, and (b) retardation in the output transition of the cell thereby resulting in breach of setup time in the later stages of flip-flops [100]. Seifert *et al* [101] presented experimental results that indicate soft error rate to a maximum possible extent of 30x is attainable in a redundancy-based approaches when a pass transistor is more vulnerable to single event upsets. The **single event upset tolerant** (SEUT) is a variant of DICE cell, designed and implemented at Intel Corporation. In SEUTs, four redundant latches are joined back-to-back and organized to form a feedback loop. DICE and SEUT differ in write logic. The provisioning of the write process in the design of SEUT is done by two latches with clock transistors connected in series. SEUT and DICE have comparable endurance and equivalent levels of mitigating the soft error rate.

Nicolaidis *et al.* proposed **Graal** [102,103,104,105] is a technique based on two phase, non-overlapping clock design style. The odd numbered latchs are clocked at $\Phi_1$ and even numbered latches are clocked at $\Phi_2$, they are 180 degrees out of phase and their duty cycle are α: (1-α) respectively. The XOR gate compares the output of D and Q of the latches and used for detecting single event transients, single event upsets and timing fault. The sufficient time is needed for an error to navigate or traverse through XOR gates, pass through OR gates and captured by the flip-flop that flags an error thereby soft error upsets gets unmasked. The single event transients that produce transient pulses at the output signals of the combinational logic circuits can be unmasked by an XOR gate such that input signal to the latch under test maintains changelessness until the flip-flop captures the error. Presumption of recovery subsystem in the micro-architecture, Graal has the capability to detect timing faults and can be operated close to minimum levels of functioning in the system. **Built-In Soft Error Resilience** (BISER) [106,107] were proposed from Intel and Stanford. We consider only the latch version of BISER. BISER Latch provides detection for SEU using redundant latch and C-element [108] proposed by Muller *et al.* and Barky *et al.*. C- Element is asynchronous circuit where the output takes the value of input only when both inputs are same. C-element with keeper protects the flip-flop from SEU by passing old value when two latches provide different value due to upsets. Keeperless C-element is called Guard gate and is much vulnerable to noise due lack of keeper [152,153]. Seifert *et al* [101] reported that SER sensitivity varies by factor of five when the minimum distance between sensitive nodes is increased by factor of 2x. C-element suppresses all SETs with longer pulse width and does not have ability to detect timing faults.

The **RAZOR - I** [109] by Ernst *et al.* flip-flop is designed to timing fault faulty with significant power reductions. The Alpha processor implemented the RAZOR - I at 180 nm technology and only 192 of 2408 flip-flops were RAZOR flip-flops. In the RAZOR-I, master flip-flop is supplemented with an additional shadow latch which captures the input at the deferred clock. The errors get detected at the outputs of XOR gate which differentiates the output of latch and flip-flop -Q by matching. It leads to a transgression in a master flip-flop parameter such as setup and hold time there by results in metastability. The asynchronous behaviour in the signals coming from latchs and flip-flops is detected with the help of metastability detector. The RAZOR-I flip-flop is upgraded with metastability detector. The refinements in timing circuits helps in detecting and correcting single event upsets. The event that is transient that lasts a duration larger than the time interval between the capturing of signals at master flip-flops and shadow latch goes undetected or unmasked to the downstream flip-flops. Shorter interval events like transients that materializes through master flip-flop signal sampling and followed by shadow latch signal sampling goes undetected or unmasked. Thus, single events transients are not detected by RAZOR -I.

The limitations of RAZOR – I were overcome by team of researchers (Blaauw *et al.* and Das *et al.*) at University of Michigan, designed and developed an enhanced variant of RAZOR -I is popularly known as **RAZOR -II** [110,111]. The redundancy, power and area overheads of RAZOR – I are high when compared to RAZOR-II. RAZOR-I does not provide safeguards to counter radiation induced errors. The challenges in the process variations does not help in designing metastability detector, a fundamental block in RAZOR – I. The RAZOR-II latch comprises of positive edge triggered latch used for master storage. The fundamental presumption in the design of RAZOR-II is that change in the latch output happens soon after the arrival of positive edge of the clock. The single event transients, single event upsets and timing faults can be detected using transition detector circuit which is supplemented to the output line of the latch. To prevent erroneous triggering, transition detector is deactivated during the negative edge of the clock after the positive edge through clock to output interval of the latch. Single event transients and single event upsets are detected by transition detector.

The **Transition Detector with Time Borrowing** (TDTB) and **Double Sampling with Time Borrowing** (DSTB) are developed by Intel and [112] Bowman *et al.* RAZOR – II and TDTB have similarity in implementation of fault tolerant designs at the latch level. The latch is upgraded with a supplement circuit named Transition detector. DSTB is a mitigation approach similar to RAZOR – I, where the master memory element is a latch with clock and is supplemented with shadow memory element called edge triggered flip-flop that stores a mirror copy. The DSTB and TDTB have similar levels of abilities to detect timing faults and single event transients. Fascinating to reflect that, a resolute metastability detection is used in RAZOR -I design to achieve maximum endurance levels [111]. However, in RAZOR- I hold-time and setup-time restraints are never met. Nonetheless, metastability is not challenge in DSTB and TDTB [112], for the reason that master memory element is a latch. TDTB do not have the ability to capture single event upsets unlike DSTB which can capture single event upsets.

**Bubble RAZOR** [113,114] architected by Fojtik *et al.* proposed a two-phase latch-based design augmented with XOR gates to detect errors, as it is similar to Graal. The published work of Bubble Razor focuses on system level error recovery in a distributed approach. The circuit is partitioned into clusters. Each cluster contains latches in the same clock phase. When latch signals error in the one cluster, it is propagated as bubble to downstream cluster. When more than one error is created, two bubble are created and receiving cluster stalls one and propagates only one bubble. By this way, two latch can handle multiple upsets propagating in the system. The first model of Bubble RAZOR is incorporated in the ARM Cortex – M core ASIC synthesized at 45nm process node size technology. A tool that translates primary flip-flop-based design to latch based design is used in the front-end design phase. The translation of flip-flop-based design to latch based design generate 8% area overhead and every flip-flop is transfigured to 3.29 latchs on average in the design. The Bubble RAZOR has a comprehensive area overhead of 21%. Bubble RAZOR technique facilitates and improves the performance by 22% under conventional process conditions at constant supply voltage. Under constant levels of performance, 54% reduction in the total energy consumed is realizable. The single event upsets are not detectable and correctable by Bubble RAZOR. Single event transients and timing faults are successfully detectable by Bubble RAZOR.

Valadimas *et al.* proposed **Error Detection Correction** (EDC) flip-flop [115], the XOR gate, master flip-flops, multiplexer, and redundant latch are the principal components of this design. The master flip-flop input and output are matched using XOR gates and its outcome is captured by the additional latch. Conceptually, EDC flip-flop maintains input for a period of interval Φ such that supplement latch and the master flip-flop capture the same data. The pulsed clock produced by specific clock circuit defers by Φ with the master clock is given to the supplement latch. Considering all the bottom-line conditions for normal functioning of EDC flip-flops, pulse width of the clock should be more than the stated admissible pulse width of the latch. EDC flip-flop can detect timing faults and single event upsets but not the single event transients.

The **Soft Error Tolerant flip-flop** (SETFF) [116] is designed to detect the SEU and has similar aspects of ED and RAZOR –II. The idea is that the flip-flop detects unexpected transitions at the output and it corrects them using the preset and clear inputs to the flip- flops. The SETFF uses a Transition detector to detect an unexpected output transition. SET provides a marginal protection to SEU because, a small window of time after clock edge, SEU cannot be detected. The TDE (Transition detector

**15**

enable) signal must be designed to cover clk to Q delay which is a longer period of time than the actual delay. A simplified version of SETFF Flip-flop has less area overhead which can only detect SEU. This SEU detect only SETFF can be used where microarchitecture correction techniques are used. The SEU detect SET FF does not provide a framework to detect or correct SET and Timing fault.

Fault tolerant designs sequential circuits are itemized in the **Table 5.**

**Table 5 Hardened designs for sequential circuits**

| Hardened designs for Sequential circuit | Area Over head (Transistor's count) | Detection | | | Correction | | | Phase latch/flip-flop |
|---|---|---|---|---|---|---|---|---|
| | | SEU | SET | Timing fault | SEU | SET | Timing fault | |
| SET FF | 53T | ✓ | | | ✓ | | | Flip-flop |
| RAZOR -I | 48T | | | ✓ | | | ✓ | Flip-flop |
| EDC | 44T | | ✓ | ✓ | | ✓ | ✓ | Flip-flop |
| DSTB | 38T | ✓ | ✓ | ✓ | | | | Single phase latch |
| BISER | 36T | | | | ✓ | ✓ | | Single phase latch |
| RAZOR II | 31T | ✓ | ✓ | ✓ | | | | Single phase latch |
| TDTB | 30T | | ✓ | ✓ | | | | Single phase latch |
| B-RAZOR | 30T | | ✓ | ✓ | | | | Two phase latches |
| SET FF (SEU Detect only) | 21T | ✓ | | | | | | Flip-flop |
| Graal | 14T | ✓ | ✓ | ✓ | | | | Two phase latches |
| Parity codes | 12T | ✓ | | | | | | Flip-flop |
| DICE | 10 T | | | | ✓ | | | Single phase latch |
| TMR | - | ✓ | ✓ | | ✓ | ✓ | | Flip-flop |

## 8 FAULT TOLERANCE METHODS FOR SRAM BASED FPGA

SRAM based FPGA devices are CMOS devices. The reconfigurability on field is the primary attribute and hallmark of SRAM based FPGA devices. The ability to reconfigure depends on the configuration setup bits stored in the configuration memory. The most susceptible regions of FPGA are configuration memory. There are two layers in the FPGA, they are logic layer and configuration layer. The interconnected input/output ports and configuration memory are the basic blocks of the configuration layer. The reconfigurability is defined by the configuration bits known as bitstream, a collection of bits that provides the sketch for a circuit in FPGA. Lookup -tables, multiplexer for control logic, and other control logic elements are realized in the configuration memory cells. Boolean logic or Boolean function are realized using lookup-tables in the configuration memory. The pass transistor or interconnection point is regulated by the bitstream housed in the configuration memory, altogether forms interconnection structure for the logic circuit. The memory cells in the configuration memory also store values for the selection lines of a multiplexer. The logic layer comprises or consists of I/O blocks, Block RAM(BRAM), configurable logic blocks, and other resources for user programs. The present state of the logic circuits is stored in registers and BRAM of the logic layer. Firstly, we discuss difference in approach for failure rate estimation for SRAM based FPGA and Digital CMOS design. Obviously two logic designs identical will have different FIT rate when implemented on SRAM based FPGA and Digital CMOS technology. Representative article in literature for fault mitigation (specifically soft error) techniques for combinational, sequential logic and configuration memory are discussed. At the end of every sub-section, important characteristics of every mitigation technique on reliability with respect to logic layer and configuration layer are highlighted in a **Table 6 and 7** respectively.

**Failure Rate Estimation in Digital CMOS vs SRAM based FPGA: -** The masking effects makes Digital CMOS circuits less vulnerable as compared to SRAM based FPGA. The soft error rate estimation and failure rate probability of a circuit differs in FPGA from Digital CMOS [118] and probable reasons are listed below.

(a) In a SRAM based FPGA, transients' events or upsets in the bitstream of the configuration memory persist or carry on up to next clock cycle. Next clock cycle has the same failure rate probability or failure rate as the previous clock cycle. The transients' events or upsets in the Digital CMOS circuits are suppressed or masked and is prevented from propagating to the downstream circuits.

(b) The logic gates in the Digital CMOS circuits are vulnerable to single event upsets that results in errors, but the routing signals. The routing signals are susceptible to transients in SRAM based FPGA unlikely in Digital CMOS circuits.

(c) Absence of electrical masking in FPGA allows the upsets in the bitstream of the configuration memory to navigate to the output signal lines of the system. Functional, logical and electrical masking makes the combinational circuit more robust against soft errors [119] in the Digital CMOS systems.

**Fault tolerant designs in logic layers: -**Shailesh Niranjan *et al.* proposed **SEU- Immune TMR** (SEU- I TMR) [120] adopts a design with single excitation circuit and state machines safeguarded by the state variables. The excitation circuits are not triplicated in SEU-I TMR. It is standard practice to have three copies of the state flip-flops, excitation circuits, and triplication of present state variables are chosen or voted so that valid present state is formed. The next state is produced using single excitation circuit in SEU-I TMR and its output is given to the triplicated state flip-flops. The voter attached to the flip-flops chooses the valid present state, unlikely the design format to choose a valid present state in a traditional TMR. The Hamming codes with a hamming distance of 2 to detect single bit errors are used to encode the state is adopted in the fault tolerant state machine proposed in the duplex architecture [121,122]. The hamming codes are used for error correction [123,124] in **explicit error correction** (EEC) method and the excitation circuit creates parity bits and state variables by employing the outputs from the non-redundant flip-flops that stores circuit state data. **Modified EEC** proposed by Shailesh Niranjan *et al.* [120], a design which has an emplacement or region occupied by error correcting logic in between excitation circuit and the state flip-flops. Error correction operation consumes less computational time in the Modified EEC when compared to EEC. Less time is due to the errors in the state variables which are non-redundant, are corrected in the error correction circuits. Protection level against the single event upsets in modified EEC and SEU-I TMR are comparable and meet the standards of TMR. ECC is incorporated in the TMR [125] has been proposed by Fernanda Gusmano de Lima *et al*.

C. Bolchini *et al.* proposed **Totally self-checking Finite state machines** (TSC FSM) [126], design procedure for hardware implementation of self-checking systems imitative of the description presented using hardware description language called VHDL. Hamming codes with fixed distance used for encoding present state and next state, heuristic based ingenious state encoding algorithms were introduced in the TSC FSM. The TSC FSM adopts Berger code encoding for outputs and checkers in their designs. The fault coverage greater than 99% is attained by using Berger codes for output encoding and in control logic unit.

Fernanda Lima *et al.* proposed **Duplication with comparison** (DWC) with **Concurrent error detection** (CED) [127], an ingenious dependable design that provides safeguards against single event upsets in SRAM based FPGA, address to two important challenges and they are (a) irreversible damage caused by single upset events in the combinational circuits, and (b) pins and energy overheads in TMR. This approach detects the transient faults that causes irreversible damages in the configurable matrix. Hotbackup, a DWC method can accomplish CED, guarantees valid or precise value at the output juxtaposed with existence of single fault is observable, indicates the capacity or ability to capture faults and rectify them on fly in 99.97% cases of fault occurrence with a latency of latency of one clock cycle before being detected by the other flip-flops.

Anurag Tiwari *et al.* proposed **Finite state Machine into synchronous embedded memory blocks** (FSM-SEMB) [128], design procedure optimized for minimizing area and power maps the FSM onto the synchronous embedded memory blocks. It is known that 14% of the total power is drained in the clock circuits, 16% in logic layer, and 60% in the interconnects present in FPGAs. In this method, encoded state bits form the data for the memory cells, memory address is function of input and present state, value stored in these memory addresses is the output and next state of the sequential circuit.

Anurag Tiwari *et al.* proposed **Enhancing reliability of finite-state machines** [129], a method that finds the unused SEMBs, are configured for FSMs of user applications in the FPGA. The parity bit are generated in the SEMBs to capture and correct single event upsets by re-inscription of the values in the SEMB forms the basis for enhancing reliability in FPGA. Re-inscription of SEMBs values is performed by maintaining an additional SEMB in the FPGA, or sourcing a write signal from the peripheral circuit. Power optimized placement of FSMs in the SEMBs in the FPGA mandates reduced number of SEMBs to be used configuring the FSM. Two power saving methods are proposed. They are (a) each state in the FSM is protected by parity bit to capture single event upsets and scrubbing from outer systems are performed to correct the errors when found, and (b) attaching parity bits for every word along with DMR enhances the detection capability to capture multiple upsets by sustaining online operations. This method has tangible implications on the reliability of SRAM based FPGA circuits with optimal power consumed.

S. Baloch *et al*. proposed **Temporal Data Sampling** (TDS) [130] technique with five edge-triggered flip-flops. There are two states or modes in which a flip-flop would operate, and they are (a) sampling mode, and (b) blocking mode. In the rising edge of the clock, flip-flops functions in the sampling mode, and in the falling edge of the clock, flip-flops functions in the blocking mode. In the sampling mode of operation, flip-flops capture the arriving data. In the blocking mode, flip-flops maintain the data and do not allow to modify them. In an interleaved regular interval of time, same data are housed in the memory elements, forms the basic principle of TDS. The samples of same data are compared and voted up on to mask the single event upsets. CLK-A, CLK-B, and CLB-C are the three independent clocks used in TDS. The main clock derives the three clocks with a phase shift of 90 degrees with 25% duty cycle to manage the single event upsets. The phase shift in the clock signals facilitates the corresponding flip-flops to store the valid data at regular intervals of time. Thereby, single event upsets induced by radiation on any clock line can be avoided or its presence is eliminated. Radiation induced transients of shorter time intervals, having an implication on the falling edge of any clock signal will not last long, because other redundant flip-flops begin their functions with respect to phase shift in the clock signals fed to them.

Thus, TDS manages single event upsets on data or clock supported by phase shift clock circuit.

Maico Cassel *et al.* proposed **Duplication with self-checking and TMR** [131], a technique that duplicates FSM. It supplements logic blocks in the FSM to capture an error. The error caused by single event upset in the FSM were managed properly by resuming the normal operation in an automated manner. The states in the FSM are encoded using one-hot method, classical method and, many other methods. Non-identical self-checking checkers were adopted for every diverse encoding style used in the FSMs. Two non-identical self-checking methods were explored in the one-hot state assignment method. In the first method, parity bits were used for investigating the present state. In the second method, FSM state transits to unused state coercing the outputs to zero, when a single event upset strike the flip-flops. This approach supplements recovery path to resume from incorrect state. Triplicated FSM operate simultaneously in TMR with most of the voters placed at the logic outputs and flip-flops [130,131]. Andrzej Krasniewski *et al.* proposed **Concurrent Error Detection for FSM** [132], in this approach synchronous embedded memory block in the FPGA are utilized for realizing FSMs. It is established that this method captures all transient or permanent faults related with one output and input of any functional block that provides incorrect output or state [104].

Kai-Chiang *et al.* proposed **Redundancy addition and removal** (RAR) [133], advocated redundant wires in the logic circuits along with logic optimization method which searches for wires that could be added or removed based on the needs of the circuit. In the automatic test pattern generation, additional wire meant for augmenting the traffic capacity is selected using the established guidelines practised for assigning wires. RAR method has very smaller cost penalty due to area overhead, because the augmented wires in the circuit always have the candidate wires for removal. New parameter to analyse the implication of the masking is presented by authors. A procedural approach-based algorithm which uses the metric to estimate the soft error rate reduction on supplementing/ dislodging redundant wires, and also helps in decision making, whether to allow/disallow changes made that are incorporated in the circuit. Total Reduction by 17.4% in the soft error rate is observed. Kai-Chiang *et al.* proposed **Selective voltage scaling** (SVS) [134], in which the gates with high supply voltage generate a comparable volume of charges that creates lesser magnitude transients which are less damaging. The low supply voltage at the gate do not generate large volume of charges that results in transients. The candidate gates that have larger implications on system soft error rate are given high supply voltage and other gates are supported with operational supply voltage. The reduction in circuit soft error rate by 33.45% for non-identical sized of transient impulses with 11.47% increase in energy consumption. Normalized power-area-delay product is 0.64% for every 1% reduction in soft error rate.

K.C. Wu *et al* proposed **Clock skew scheduling** (CSS) [135], in which the probability of the sampling trivial transient impulses is drastically reduced, when an incoming clock signal time period is scheduled accordingly thereby masking frequently by using latching window is increased. The particle strikes a clock cycle in the single event upsets thereby rising the likelihood of timing masking through by adopting proper scheduling. The **Built-in self-test** (BIST) [136] is approach when circuit performs a self-testing. BIST is a collection of test procedures used for testing memories and embedded logic blocks in FPGA. All the operations in FPGA are regulated by the BIST controller. This approach does not enhance reliability by high margin. Aiman *et al* advocated **Redundant equivalent states** [137], an innovative concept that enhances the reliability of the sequential logic circuit by supplementing redundant states to the existing vulnerable states which has greater likelihood of particle strike. The states that are vulnerable to particle strike are safeguarded by redundant states thereby reliability of the logic circuit is enhanced when an upset occurs in one among these states or transiting to it. The augmented FSM with redundant states is comparable to the original FSM, such that redundant states have same output and the next state as of safeguarded states.

**Fault tolerant designs in configuration layer: -** Srinivasan *et al.* proposed **Asymmetric SRAM** (ASRAM) **and Refreshing SRAM** (RSRAM) [138] to improve reliability of configuration memory. Majority of the bits in the configuration memory are set to zero. The ASRAM – 0 cell are suitable to store configuration bits thereby reduce leakage in the configuration memory. The configuration memory has a leakage of 38% when realized at 90nm is observed by evaluating the experimental results. Meanwhile, reduction in the vulnerability of the configuration bits (housing zeros) that employ ASRAM-0 cells towards soft error is evident. Usage of ASRAM-0 cells diminishes the FIT rate by 50% when compared to the traditional SRAM cells. Balkaran et al proposed **RSRAM0** [139], an asymmetric cell similar to ASRAM-0 cells are permanent-0 cells. The RSRAM-0 cells are synthesized at 70nm process node technology. If all the cells of the configuration memory are housing zeros, then RSRAM-0 cells will make sure FIT of '0' for the entire memory. Similarly, ASRAM-0 cells usage will result in a FIT less than conventional SRAM. Maximal FIT is observed for housing 'ones' in ASRAM-0 cells. The FIT of the conventional SRAM cells housing 0 and 1 is greater than RSRAM-0 cells.

N. Rollins *et al.* proposed **TMR coupled with scrubbing** [140], is a technique in which upsets are masked or corrected by using TMR during the scrubbing interval as result of that improvements in the reliability is observed. The vulnerability of the configuration memory bits can be decreased by employing feedback TMR designs and TMR with three voter system. The TMR in the feedback path has shown more efficacy in masking single event upsets. For example, in case of 8-bit counter, TMR in the feedback path and global clock triplicated has decreased the number of vulnerable bits in the configuration memory to zero, also obliterating issues related to re-synchronization. In the few design alternatives, voters and counters are stuffed into single LUT in TMR based feedback circuits. A high-cost dependable system has been realized by the researchers. Zarandi *et al.* proposed **Hard-wired switch module structure** [141] proposed hybrid switch module comprising of (a) configurable structure, and (b) hardwired structure in FPGAs, to tolerate single event upsets. Enhancement in the reliability of the interconnects in the switch module by 30% is observed in programmable structure-based switch module, while vulnerability to single event upsets is reduced by 50% in hardwired based switch module. Melanie Berg *et al.* proposed **Scrubbing and Partial reconfiguration** [142] proposed execution of scrubbing after the write operation in the configuration memory. Scrubbing can be realized in a FPGA using internal approaches (software or hardware) or external approaches. Innovative approach proposed by Heiner *et al* [90] named "partial reconfiguration" is implemented along with scrubbing in the configuration memory. Unceasingly the whole FPGA is scrubbed by self-scrubber residing in smaller region simultaneously reorganizing a region in the design.

Feng *et al.* proposed **In place reconfiguration** (IPR) [143], assumes randomized one/single fault model. IPR proposes an algorithm that executes logic conversions while retaining LUT functions and topology of the logic circuits. IPR improves MTTF by 1.94x and marginally decrease fault rate by 48% for the invariant area and performance. IPR along with resilient logic transformations algorithms decreases the fault rate by 49% and improves MTTF by 2.4x with reduced area by 19% with constant performance. Lee *et al* proposed **In-place decomposition** (IPD) [144] with a presumption of stochastic single fault model for configuration bits affected by faults inflicted by single event faults. IPD breaks down a Boolean function for a logic block in to two Boolean functions, realized by utilizing LUT with dual outputs in the logic blocks, connected to carry chain in the logic blocks thereby minimizing fault rate at the chip-level. Ebrahimi *et al.* proposed **Mitigate bridging and short faults** [145], the principal idea of this method is to decrease the number of SRAM cells in the switch box module thereby reducing the vulnerability against single event upsets. The SRAM cells used in the switch box is reduced by 81.4% when compared to traditional SRAM. It is remarkable to note that fall in the SRAM bits count in the switch module is proportional to declining rate of single event upsets. Thus, dependability of the switch box is enhanced by 18.6%. Zhe Feng *et al.* proposed **In-place x-filling** (IPF) [146], is method that gives consideration to single event upsets in the interconnect bitstream and LUT bitstream. IPF provides reasonable reliability when compared to other fault mitigation techniques for LUTs and Interconnects. IPF enhances the system reliability by masking single event upsets in the interconnects and LUTs. IPD fails in safeguarding the interconnects against the single event upsets. Chip level MTTF of 7% enhancement is attained in IPD is because of 4x enhancement in LUT MTTF. 21% enhancement in MTTF at chip level is attainable by using IPF. IPF is resilient to soft errors in the fan-in cones, thereby diminishes single event upsets in fan-out cones too. MTTF enhancement up to 52.6% at the chip level is attainable in IPF, as error transmission is similar in interconnects and LUTs. IPD is slower by 128x than IPF. Hence IPF is reliable scalable and suitable for marketable FPGAs.

Ebrahimi *et al.* proposed **Modified switch box** [147] consisting of four SRAM cells connected to six switches using single decoder-based switch box. In the conventional switch box, six switches are directly connected to six SRAM cells. In this modified switch box, only 75% of the SRAM cells are used when compared to the conventional switch box thereby reducing the vulnerability of the soft errors in the switch box. In the absence of redundancy, this architecture enhances the reliability of the switch box is proved. Smaller latency overheads and area overheads with improved reliability, the switch box would be suitable design for use in industry. Uros Legat *et al.* proposed **SEU Recovery Mechanism** [148], in this mechanism Internal configuration access ports (ICAP) are used for error – recovery operations to neutralize the effects of errors in the configuration memory. The recovery procedure resides in a little region of the FPGA, at the same time remaining regions of the configuration memory are used for user programs. ICAP through which bitstream is read or written. The 12 parity bits are encoded in the bitstreams of virtex-4 and virtex-5

FPGAs. The hamming codes are used for ECC. The syndrome value determines the error in the bitstream consisting of 1300 bits of data. For every read in virtex-4 and virtex-5 FPGAs, ECC is used to generate the syndrome value, to detect errors in the configuration memory.

Table 6. Fault tolerant designs in logic layer

| Fault tolerant Designs | Important Characteristics |
|---|---|
| BIST | Self-test procedures for memories and logic blocks, and improves reliability |
| Clock skew scheduling | Masks timing faults, schedules clocks, and improves reliability |
| Concurrent Error Detection for FSM | Usage of SEMBs, and improves reliability |
| DWC with CED | Reduces pin overheads, and rectify on fly fault occurrence in 99.97% cases |
| Duplication with self-checking and TMR | Redundancy, masks SEU, usage of self-checkers, and enhance reliability |
| Enhanced reliability of finite-state machines | Parity bits, Power optimized placement of FSM |
| Redundancy addition and removal | Redundancy (wires), Reduction by 17.4% in the soft error rate is observed |
| Redundant equivalent states | Redundant states to protect vulnerable states |
| SEU- I TMR and Modified ECC | Single excitation circuits, and hamming codes |
| SEMB | Optimized for area and power usage |
| Selective voltage scaling | reduction in circuit soft error rate by 33.45%, and. Normalized power-area-delay product is 0.64% for every 1% reduction in soft error rate |
| Totally self-checking Finite state machines | fault coverage greater than 99% is attained by using Berger codes |
| Temporal Data Sampling | Temporal redundancy with three clocks with varying phases |
| TMR with ECC | error correcting logic in between excitation circuit and the state flip-flops |

Table 7 Fault tolerant designs in configuration layer

| Fault tolerant designs | Important characteristics |
|---|---|
| Asymmetric SRAM and Refreshing SRAM | Usage of ASRAM and RSRAM -0 cells, FIT reduced drastically |
| Hard-wired switch module structure | Enhancement in the reliability of the interconnects in the switch module by 30% |
| IPR | IPR improves MTTF by 1.94x and marginally decrease fault rate by 48%, |
| IPD | Minimizes fault rate at chip level |
| IPF | 21% enhancement in MTTF at chip level is attainable by using IPF |
| Mitigate bridging and short faults | Dependability of the switch box is enhanced by 18.6% |
| Modified switch box | only 75% of the SRAM cells are used when compared to the conventional switch box, thereby improves reliability |
| TMR coupled with scrubbing | Scrubbing with TMR in feedback paths |
| Scrubbing and Partial reconfiguration | Scrubbing can be done internally or externally |
| Single event upset Recovery Mechanism | Hamming codes for error correction |

## 9 FAULT TOLERANCE METHODS AT MICRO-ARCHITECTURAL LEVEL

Conceptually, the superscalar data path is used for detecting faults either by replicating a stage in a pipeline or re-executing a portion of a program during idle cycles of the pipeline. The replicating structures at pipeline level can also be called micro-architectural redundancy. Re-execution of instructions to detect is called as time redundancy. The idea of replicating execution units was advocated by Parashar *et al.* [155] and Ray *et al.* [154]. The instruction is dispatched to execution units from renaming unit and the results are compared by hardware structures. If the results do not match, occurrence of hard faults in execution units can be inferred. Few representative error detection mechanisms are discussed in the following section. We end this section with summary of error detection mechanisms listed in the **Table 8**.

**Fault mitigation techniques: -** Gomma *et al.* proposed **Partial explicit redundancy and implicit redundancy through reuse** (PER-IRTR) [156] is a technique that uses explicit redundancy, that is the main thread uses the computing space which is not utilised by the replicated instructions. Typically, the main computing space is utilized by replicated instruction racing against the main thread in the processors. Explicit redundancy is provided only in low –ILP phase. For example, during cache miss, to safeguard against the soft errors, the instructions execution is replicated. During the high ILP, there is no soft error detection and no performance degradation exist, as thread achieve peak execution. Instruction reuse IR is the conceptually re-executing the instruction many times with same input. The reuse buffer used to the input and output values. In Dynamic IR [157], it avoids the re-execution of instruction if the output matched for two instances of input. Accordingly, reuse buffer is updated. Samuel Williams *et al* proposed **Dual instruction execution – reuse buffer** (DIE-IRB) [158], is technique where the results of the main thread and later thread for capturing the fault. In IRTR, updating buffer is carried out by the main thread by not matching. The output of the subsequent instructions is compared with reuse buffer values. A strategy adapted for sub maximum coverage for the given work load with no redundant threads. The soft error rate decreases in PER+IRTR, PER and IRTR by 56%, 44% and 22% respectively.

Meixner *et al.* proposed **ARGUS** [198] is a technique that checks control flow, computation, and dataflow and memory correctness at run time in simple cores. **Dynamic Dataflow Verification** (DDFV) [159] is a technique that adopts dataflow and control flow checker developed for superscalar core is used in the ARGUS. ARGUS uses Data and control signature DCS single signature are inserted in every basic block along with the predecessor's one, unlike DDFV which inserts only one signature or compute for a basic block. The permanent faults and transient faults are detected by ARGUS with lesser implications on performance of the simple cores. It has area overhead of 17 % and performance overhead of 3.2-3.9% with fault coverage of 98% for permanent and transient faults. Latency to detect transient faults is very less, an important ability and merit that the ARGUS possess.

**Table 8 Fault tolerant designs at micro-architectural level**

| Concept/Designs | Detection mechanism | Performance over head | Detection latency | Protects | Faults target |
|---|---|---|---|---|---|
| **PER – IRTR** | Re -execution | 2% | bounded | core | Transient faults |
| **ARGUS** | Monitoring invariants | 4% | low | core | Permanent and Transient faults |
| **BulletProof** | BIST | 5-25% | bounded | core | Permanent faults |
| **SHAREC** | Re-execution in idle cycles | Medium | bounded | Backend of core | Transient faults |
| **SeIR** | AVF estimation | Low | Bounded | Backend of core | Transient faults |

Shyam *et al.* proposed **BulletProof** [160] is a technique that protects pipeline and on chip memory. To detect and isolate permanent faults, distributed BIST is employed in the pipeline of BulletProof. BulletProof do not have the ability to capture soft errors. It uses 4-wide Very long instruction word (VLIW) processor with separate cache for data and instruction. A checker unit is connected with every stage of pipeline like instruction decoder, arithmetic logic unit (ALU), register file and cache. The ALU is checked using 9-bit mini ALU. It has 89% coverage for silicon defects with small area cost 5.8%. Influence on performance is less, a significant gain in this method. Smolens *et al.* proposed **SHared REsource Checker** (SHAREC) [161] is a technique that executes instruction asymmetrically like IRTR. The performance penalty in multithreading approaches is due to sharing of issue queue, functional unit and reorder buffer (ROB). SHAREC moves instruction in program order from ROB into small in-order queue for re-execution. The functional unit is common to in-order issue queue and standard issue queue. The replicated instruction or wait for re-execute instruction are executed in the functional units during the idle cycles of original stream of instructions. Redundant Multithreading Techniques (RMT) provides safeguards encompass the frontend, unlike SHREC offers safeguards for core backend. Xavier Vera *et al.* proposed **Selective Replication** (SeIR) [162] is a technique that replicates instruction in issue queue by estimating AVF and avoids re- directing the instruction from commit phase. SeIR reduces pressure on ROB unlike SHAREC.

## 10 FAULT TOLERANCE METHODS AT ARCHITECTURAL LEVEL

Multi-core processor architecture consists of CPU cores, cache memory, register file, interconnection logic, and main memory. In processor die, major area is occupied by memories which are protected by error correcting codes (ECC). Fault tolerant designs are provided to protect the remaining portion of the die covering CPU and memory hierarchy control logic. We present a brief survey of fault mitigation techniques at processor core level that improves the reliability of the multi-core system. A brief on reconfiguration approaches, core salvaging approaches and recovery techniques are also presented. Current state of the art error detection techniques can be grouped in five major classes as: fault detection by redundant execution approaches that make use of built-in duplication at (a) processor cores level, (b) thread level in a multi-core processor architecture, and (c) instruction level (Software based redundancy techniques). Other two approaches are (d) BIST methods provide self-test routines for processors conventionally used for manufacturing testing (e) Fault symptoms detection methods detect symptoms that could possibly result in transient faults.

**(a) Error detection by redundancy at core level: -** Classical n-modular redundancy approaches have been used for detecting the errors. Triple Modular Redundancy (TMR) and Dual Modular Redundancy (DMR) methods are like special case of classical n-modular redundancy systems. Modern processors running in lock step mode and comparing the results in every cycle is challenging due to delays in signal propagation in source synchronous buses [163]. **Tight lock-step redundant execution: -** In these systems, two cores execute the same instruction and compare the outputs cycle by cycle to detect errors. The architectural state of the two cores must be same and is achievable by identical initialization. Lock-stepped machines are not suitable for commercial market except for mission critical systems [164,165]. The first marketable dependable servers were available from Tandem Computers Inc, adopted DMR at core level. Cores were tightly synchronized for instruction execution. Later Tandem became HP Non-stop enterprise system. The system consisted of dual-bus interconnect connecting 2 to 16 processors. As soon as the fault is detected, the processor stopped working. Hence, they are known as fail-stop machines. The parity modules and self-test modules formed the basis for fault tolerant computing in Tandem processors. **Loose lock-step redundant execution: -** In these machines, core executes the same instruction but they are not synchronized. The input not being coherent is a problem or challenge to be addressed in loose lock stepping among cores. The trailing core may read a value, which the leading core may feed a different value during runtime. **Non-Stop Advanced Architecture** (NSAA) [166] proposed by Bernick et al. and **configurable isolation** [167] proposed by Aggarwal et al. falls under this approach. The architecture of NSAA has three groups of 4-way Itanium server processors. Core from each group form a slice, each executing the same program. But they separate address space. NSAA prevents fault propagating beyond sockets and pairs process at the socket level.

Smolens et al. proposed **Fingerprinting** [168] is a technique in which outputs are compared after the register updates and it increases the bandwidth and is called as full state comparison. The Cyclic redundancy code is used to generate a hash known as fingerprint, is the compressed signature of branch predictors, load and store address, and register updates in the core architecture. Configurable isolation [167], is a technique in which the implication of fault containment at memory level are studied. The fault zones are configurable and provide performance degradation at high failure rate. The DMR or TMR or even fault isolation domains of 4 can be configured at runtime. The reconfiguration is done when a fault gets detected in the system. Failed cores are ignored in one region and the non-faulty enduring cores continue to be deployable. After 10 years of operation, Configurable isolation continues to perform at 60% of the early lifetime phase. Soft errors and hard errors are detected in this technique. LaFrieda et al. proposed **Dynamic Core Coupling** (DCC) [170], is a technique that permits communication between replicated threads on different cores to use the common system bus of the shared memory. DCC prevents static coupling of the processor cores with 3-5% performance overheads, but allows cores to compare their results at runtime in a DMR arrangement. Both hard errors and soft errors are detected and corrected by DCC. The Backward Error Recovery (BER) is used to resume normal operations in DCC when it encounters soft errors and hard errors.

Austin et al. proposed **Dynamic Instruction Verification Architecture** (DIVA) [171] is a technique that adopts two cores, and they are primary core and checker core, primary core executes all instructions excluding load and stores. The primary core executes in out of order and the checker core verifies the results from primary core in order fashion. The checker core has been added to commit phase of the processor pipeline and has area overhead of 6% to ALPHA21264 processor. The instructions, values including the speculative results from reorder buffer are moved to checker core. DIVA exploits asymmetric execution, thereby eliminating the dependencies between the instructions. Similarly, Beta core solution (BCS) [172] incorporate a minimal in-order core for re-execution of bundle of instruction by generating signatures. BCS do not detect faults and only can identify bugs. Smolens et al. proposed Reunion [173] is a fingerprint technique which has two cores - a vocal core and a mute core. On Chip multi-processors with two cores execute the replicated identical threads. The inputs to the identical cores are same and ensured by using rollback recovery-based protocol for repeating the execution. rollback recovery-based protocol is called when two cores' results differ in their results. The detection latency is very less, but performance overhead is 9% and 8% on commercial and scientific workload and is 5-6% from relaxed input replication.

Sundaramoorthy et al. proposed **Slipstream** [174] have similarity with DIVA, but two cores (leader – follower) have identical microarchitecture. In DIVA two cores are not identical, unlike Slipstream has 100% hardware overhead and has 12% speedup. The weak instructions are captured and disregarded by the leader core in Slipstream. The weak instructions satisfy the conditions like (a) computes the branch conditions of predictable branches (b) dynamically dead instructions (c) produces the same value. Unlike Slipstream, **Decoupled Performance Correctness Architecture** (DPCA) [175] proposed by Garg et al. assesses results of the branch instructions of the program while executing in a leader core and prefetches relevant instructions in the cache memory for feeding to the trailing core. The prefetched instructions increases the pace of execution of the entire program in the trailing core as the results of the branch instruction are already known and kept in the cache memory. **Paceline** [176] proposed by Greskamp et al. is conceptually comparable to Slipstream and DPCA, and deliver enhancement in performance. In this technique, the leading core is operated at higher frequency when compared to trailing core. The outcome of asynchronous operating frequency between cores facilitates timing speculation thereby trailing core can be used to capture errors. Greskamp *et al.* and Torellas *et al.* [176] advocated that with no errors, the operating frequency can be as high as 1.3 times the standard frequency. The **URISC++** [177], proposes a co-processor to TigerMIPS which executes only one instruction repeatedly emulating the critical instructions and verify the result with main processor to detect permanent fault. Venkatesha *et al* proposed 32-bit One instruction core (OIC) [88] that supports main core by emulating the faulty instruction by repeatedly executing one instruction. 32-bit OIC is fault tolerant core with less power and area overhead.

Subramanyan *et al.* proposed **Redundant Execution using Simple Execution Assistance** (RESEA) [178,179], is energy efficient technique that supports trailing cores in executions. The leading core transfers the values of branch and load instructions to the trailing cores, conceptually identical to the Simultaneous Redundant Threading [184]. The RESEA optimizes the energy consumption in the trailing core. Less than 1% performance overhead with energy consumption of 1.34x times more than the baseline executions. Moreover, energy consumption is reduced by 1.26x times to that of baseline execution with the introduction of early-write optimization in this technique. RESEA transfers single value from the leading core to trailing core in the majority of every two clock cycles. High interconnect bandwidth is the major disadvantage of RESEA.

**Table 9. Fault tolerant designs at core level**

| Concept | Performance cost | Detection latency | Protects | Source of failure | Detection coverage | Area overhead |
|---|---|---|---|---|---|---|
| **Lock-stepping** | 1.5 – 2x | Cycle by cycle detection latency | core | Transient and permanent faults | 100% | 100% hardware cost in DMR based system |
| **Fingerprinting** | Low | Very high | (State) core | Transient and permanent faults | 100% | <100% |
| **DCC** | 3%-5% | Cycle by cycle detection latency | core | Transient and permanent faults | 100% | 100% + 64-entry age table in each core to support master-slave consistency |
| **DIVA** | Low (5%) | Low | Back end of core | Transient and permanent faults | 100% | 6% |
| **Reunion** | 5%-9% | Low | core | Transient and permanent faults | Better than RMT | 100% |
| **Slipstream** | 12% speedup | Deterministic | core | Transient and permanent faults | Very high | High |
| **RESEA** | 1% | Deterministic | core | Transient and permanent faults | Less than 100% | Low |
| **RECVF** | 1% | Deterministic | core | Transient and permanent faults | Less than 100% | Low |

Subramanyan *et al.* proposed **Redundant Execution using Critical Value Forwarding** (RECVF) [180] is a technique that transfers the outcomes of the instruction executions identified along the critical path from leading core to the trailing core. This breaks the cable of data dependence in the trailing cores. Thus, it enhances the performance. RECVF transfers 33% of the regular transfers to the trailing core, thereby it reduces the traffic on the interconnects. The energy overhead of 1.26x times the baseline execution and performance overhead of 1% is observed in this approach. RECVF has become popular because of lesser interconnect bandwidth used and adopted in network on chips (NOC).

The concepts of fault tolerant designs at core level are itemized in **Table 9**.

**(b) Error detection by redundancy at thread level: -** The design of multithreading techniques to detect faults are less complex as compared to core-based techniques. The parameters like clock skew, propagation delay in buses do not affect threads communicating intermediate results. Threading provides reliability with less hardware cost. In 1995, IBM G5 [181] was the first commercial machine to incorporate multithreading by running two pipelines simultaneous. One extra hardware stage was added to verify the outcomes of the two pipelines. A lenient approach in redundant multithreading (RMT) is adopted with loose lock stepping or without it [182]. Rotenberg *et al.* proposed **Active stream/Redundant stream Simultaneous Multithreading** (AR-SMT) [183] is a technique employs two threads namely 'A' and 'R'. Both the threads execute the same application. The 'A' thread executes before 'R' thread in tens of cycles. The instruction is committed only if the results match. This approach detects transient faults and corrects them with 10-30% overhead. Reinhardt et al proposed **Simultaneous and**

**Redundant Threaded** (SRT) [184] is a technique that extends the idea of AR-SMT by dynamically scheduling the instruction from redundant threads improves performance by 16% than lock stepping. The performance overhead in SRT is about 20-30% when compared to processor with no fault tolerance. It consumes energy of 1.5-1.6x times more that of the baseline executions. Unlike AR-SMT, register file comes within the sphere of replication. Special buffers first in first out are used to transfer value and addresses from leading core to trailing core. Vijaykumar *et al.* [185] augments the SRT with soft error recovery capabilities to called SRTR. It has very high performance and power costs. Mukherjee *et al.* proposed **Chip level multithreading** [186]**,** RMT for single SMT Processor is implemented on Chip Multiprocessor (CMP) is termed as Chip level multithreading (CRT) provides a 13% better performance than lock stepped cores with detection latency of 10 cycles or more.

Gomaa *et al.* proposes **Chip-level Redundantly Threaded multiprocessor with Recovery** (CRTR) [187] is a technique that supplements CRT for identifying the soft errors. The detection latency is very high and is of 30 cycles occupying large chuck in inter-processor bandwidth. **Dead and dependence-based checking elision** (DDBCE) in CRTR do not match the results of dynamically dead instructions thereby reduces bandwidth meant for comparison and considers only instructions occurring at the tail of data dependence list for comparison. CRTR ensures register death masking faults that does pollute the results later. Parashar *et al.* proposed **Slice-Based Locality Exploitation for Efficient Redundant Multithreading** (SlicK**)** [188] is a technique that introduces collection of predictors makes effort to validate the results of the leading thread with out executing them again. SlicK depends on branch predictors. The trailing threads re-executes the instructions that has evaded the predictors validation. Sumeet Kumar *et al* proposed Speculative **Instruction Validation** (SpecIV) [189] supplements the basic principle of value prediction to all instructions in the leading thread. SpecIV incurs large overhead in area and is a not highly reliable solution.

The concepts of fault tolerant designs at thread level are itemized in **Table 10**.

**Table 10. Fault tolerant designs at thread level**

| Concept | Performance cost | Detection latency | Protects | Sources of failure | Detection coverage | Area overhead |
|---|---|---|---|---|---|---|
| **AR – SMT** | Very high | Hundreds of cycles | core | Transient faults | <100% | >2x |
| **SRT** | High | Unbounded | core | Transient faults | <100% | >2x |
| **SRTR** | Very high | Hundreds of cycles | core | Transient faults | <100% | x |
| **CRT** | CRT: achieves 13% better performance than a dual lockstep CPU | Unbounded | core | Transient and permanent faults | <100% | >2x |
| **CRTR** | Very high | 30 cycles | core | Transient and permanent faults | <100% | Very high |
| **SlicK** | Medium | Unbounded | core | Transient faults | <100% | Very high |
| **SpecIV** | Medium | Unbounded | core | Transient faults | <100% | Very high |

**(c) Software-based redundancy techniques: -** In order to decrease the hardware cost, time redundancy approaches gained momentum. Conceptually, computations are repeated to detect faults by comparing the results. **Signatured Instruction Streams** (SIS) [190], **Path Signature Analysis** (PSA) [191] and **Continuous Signature Monitoring** (CSM) [192] are hardware – software amalgamated approaches to detect the errors in the control flow at fetch and decode logic. These methods have very high detection latency and instruction can change the architecture state before they graduate. Oh *et al*. proposed **Control Flow Checking by Software Signatures** (CFCSS) [193] is a software-based techniques that performs control flow checking. The CFCSS ensures proper transfer of control to the correct descendent basic block. But no assurance is given by CFCSS that exact path of a predicated branch is chosen at runtime. The CFCSS is not a suitable approach to perform recovery in control flow for superscalar processors. Oh *et al.* proposed **Error Detection by Duplicated Instruction** (EDDI) [194] is a technique that advocates time redundancy in a software-based implementation that replicates instruction at compile time thereby resulting in 100% performance overhead. Huge memory space overhead since all the instructions are duplicated in EDDI. Reis *et al.* proposed **SoftWare Implemented Fault Tolerance** (SWIFT) [195,196] is an improved version of EDDI. The single-core and multi-core systems executing single thread applications can use EDDI and SWIFT to enhance the dependability of the systems. The improved version of control flow mechanism (EDDI+ECC+ECFE) is adopted in SWIFT thereby reduces performance overheads. SWIFT do not have branch validating code, its usage reduces performance. Only soft errors are detected by SWIFT. In SWIFT, memory is safeguarded by codes is presumed, and register files are duplicated for two chains of instructions execution. CFCSS

**24**

forms the basis for SWIFT implementation and it ensures valid directions for branch instructions are taken.

Reis *et al.* proposed **CompileR Assisted Fault Tolerance** (CRAFT) [197] is a makes necessary changes to address the limitations of SWIFT by implementing hardware buffers to enhance the dependability. The separate buffers for store and load instruction are implemented to store addresses and data to write (or read). The redundant store or load instruction checks the entries in the buffer and validates the same. After validation, buffer commits the value to memory. Wang *et al.* proposed **Software-based Redundant Multi-Threading** (SRMT) [199] is a technique that has the ability to capture only soft errors. The replicated threads created at compile time in SRMT executes simultaneously on CMP. These threads use special memory space for inter-thread communication. Chang *et al* proposed SWIFT-R [200] is a technique that adopts TMR before load/store instructions along with triplicated programs for recovery in the software systems. The program executions are duplicated in TRUMP **(Triple Redundancy Multiplication Protection),** and one copy uses AN-codes. Both soft errors and hard errors are detected in TRUMP.

The concepts on software redundancy-based fault tolerant designs are itemized in the **Table 11**.

**Table 11 Software based fault tolerant systems**

| Concept | Performance cost | Detection latency | Protects | Sources of failure | Detection coverage | Recovery |
|---|---|---|---|---|---|---|
| **SIS, CSA, and CSM** | Low | Unbounded | Control flow logic | Transient and permanent fault | Control flow errors | No |
| **CFCSS** | Low | Unbounded | Control flow logic | Transient and Permanent fault | Control flow errors | No |
| **EDDI** | >150% | Low and unbounded | core | Transient fault | 98.5% | No |
| **SWIFT** | Very high | Low and unbounded | core | Transient fault | <100% | No |
| **CRAFT** | Very high | Low and unbounded | core | Transient fault | <100% | No |
| **SRMT** | Very high | Unbounded | core | Transient and permanent fault | <100% | No |
| **SWIFT- R and TRUMP** | Extremely high | Unbounded | core | Transient and permanent fault | <100% | Yes |

**(d) Built in Self-Test Methods: -** Psarakis *et al.* proposed **Software based BIST** (SBST) [201] is a technique that generates test patterns in the processor using its instruction. In SBST applied to bus based CMPs [202][13] is proposed by Apostolakis *et al.* is a technique in which the uniprocessor test programs are deployed on all cores for parallel execution, thereby reduces the total test execution time with fault coverage of 91% for permanent faults. The SBST protracted to multithreading is **Multi-Threaded SBST** (MT-SBST) [203]. MT-SBST proposed by Foutris *et al.* is a technique that improves the self-test time at processor level by 6x and at core level by 3.6x when compared to single-threaded application. MT-SBST has a fault coverage for stuck-at fault of about 88% at chip-level and 91% at functional unit's level. The access control extension framework [204] can access the state and control of the microprocessor by periodically suspending execution and testing time varies between 5% and 25% of the system time. The concepts for BIST are itemized in the **Table 12.**

**Table 12 BIST based fault tolerant designs**

| Concept | Performance overhead | Detection latency | Faults target | Fault coverage |
|---|---|---|---|---|
| **SBST** | Frequency of self-test execution | Test period | Permanent faults | 90% |
| **MT- SBIST** | Frequency of self-test execution | Test period | Permanent faults | 90% |

**(e) Fault symptoms detection-based fault tolerant systems: -** The unpredictable behaviour is the symptom of fault in a software system is captured by continuous observation. The levels of manifestations of faults detection methods or approaches can be broadly divided into three major divisions and they are (a) OS abortion or terminations, application malfunctioning, dangerous hardware traps etc, are captured by techniques such as **SoftWare Anomaly Treatment** (SWAT) [207] ; (b) invariants at bit level, comparison of history of values with current ones, beyond range or value limits or out-of-range values etc, are detected by techniques such as **Perturbation-based Fault Screening** (PBFS) [205]; page faults, exceptions, cache misses etc, at the micro-architectural level are captured by techniques such as **Restore** [206]. 95% of the hard errors are detected in SWAT. In Restore, experimental result indicates 2x rise in mean time between failures when compared against the standard pipeline with negligible hardware and performance overheads. The errors in the latches of the pipeline are eliminated or captured by Restore. **Trace based fault diagnosis** (TBFD) [208] proposed by Man-Lap Li *et al.* manipulates the existence of non-faulty and faulty cores in the system to rerun and match the results of non-faulty and faulty cores. TBFD effectively identify 98% of faults in the faulty components activating small-grained overhaul. **Multithreaded-SWAT** (m-SWAT) [209] proposed by Sastry Hari *et al.* applies discerning TMR for identifying faults. The m-SWAT effectively detects 95% of the permanent faults, and 100% of the faults are captured that evades to non-faulty cores.

The concepts of symptoms-based fault tolerant designs are summarized and itemized in **Table.13**

Table 13 Fault symptoms detection-based fault tolerant software systems

| Concept | Performance overhead | Area overhead | Detection coverage | Detection latency |
|---|---|---|---|---|
| **PBFS** | low | Low | <100% | Unbounded |
| **Restore** | Low | Low | <100% | Unbounded |
| **SWAT** | 5-15% | Low | 95% of permanent faults | Unbounded |
| **TBFD** | 5-15% | Low | <100% | Unbounded |
| **m-SWAT** | 5-!55 | Low | <100% | Unbounded |

**Reconfiguration and repair approaches: -** Shantanu Gupta *et al.* proposed **StageNet Fabric** [210] is a multi-core system that has pipeline stages which can be re-arranged or re-configurable to isolate faulty stages. The cores share the pipeline stages can be perceived as the fine-grained redundancy that is built-in into system. To tolerate permanent faults, StageNet Fabric performs reconfiguration at module level, pipeline stage level and core level. Their study shows MTTF gains are high at module level and pipeline stage level reconfiguration. Coarse grain re-configurability is a simpler technique to implement, give poorest returns in terms of lifetime extension with increasing defects. For embedded benchmarks, the throughput is 4x times the baseline after 7 years. After 12 years, there are 12 failed modules in the system. Meixner *et al.* proposed **Detouring** [211] is a repair technique for errors in in-order cores used for constructing parallel machines. The software is reconfigured to use non-faulty cores or regions in the micro-architecture by retaining its original functions. No performance or hardware cost is incurred in Detouring when all cores are non-faulty. Similarly, permanent fault masking is done on array using self-repairing array [212] without calling DIVA [198] recovery routines. Pellegrini *et al* proposed VIPER [213] is a re-configurable technique that permits control logic distributed across hardware clusters which forms the execution engine for instructions. VIPER suffers from graceful degradation of performance in absence of single point of failures. Instructions are grouped to form bundles which are executed completely in a hardware cluster. At runtime, VIPER selects hardware clusters that can join any pipeline virtually. In presence of permanent faults, VIPER provides better performance.

**Micro-architectural core salvaging: -** The salvaging approaches capitalize on intrinsic redundancy by slating the processes on standby resources and disables the faulty portions of the pipeline. Schuchman *et al.* proposed **Rescue** [214] is repair and re-configuration technique for out-of-order architectures. Speculative execution and enhanced performance are the important characteristic of Rescue. In order to improve yield and facilitate deterioration in performance gracefully, components which are redundant and non-essential are deactivated. The yield accustomed to instruction throughput is improved by Rescue against core sparing approaches by 22% at 18nm, and 12% at 32nm process node technology. Romanescu *et al.* proposed **Core Cannibalization Architecture** (CCA) [215] adopts TMR and DMR for cores that is more advantageous to multi-core chips. CCA ensures sufficient number of functional cores to be available for TMR and DMR modes of operations. CCA for 4-core chips and 3-core chips incurs 64% and 63% respectively, increase in cumulative performance. Jayanth Srinivasan *et al.* proposed structural duplication [216] approaches which suffers from performance degradation. Structural duplication improves reliability by 3.17 times the base value for 2.25 times the base cost.

**Architectural core salvaging: -** Michael Powel *et al.* [217], proposes hardware-based thread migration to another core that can execute the operation from a core that is not able to execute such that instruction set architecture of both cores are same. This approach considers only hard faults on homogeneous cores with very high overhead area and they do not provide solutions for thread migration in multiple ISA multi-core system.

**Error recovery and repair techniques: -** Fault is an apparent form of an error in the system. Error recovery can be performed in two different ways and they are (a) forward error recovery (FER) (2) backward error recovery (BER). The correction and identification of errors in absence of changing or transiting to predecessor valid state is basic principle followed in FER to re-construct state by maintaining redundant state which is free of errors. TMR systems, DMR systems, pair-and-spare systems, and Marathon Endurance server used FER to resume normal operations from errors in the hardware. DMR systems is adopted for recovery in Stratus ftserver [218]. BER has low hardware overhead when compared to FER. Backward Error Recovery (BER) saves checkpoint (or state information) at regular intervals of time. If a fault is captured, then BER moves backward or rollbacks to verified checkpoint. Sorin *et al.* proposed **SafetyNet** [219] adopts gradual acquiring of data and uses checkpoints locally. Milos Prvulovic *et al.* proposed **Revive** [220] is a technique that adopts directory-based approach to acquire or log the changes in the memory. Revive uses global checkpointing and can tolerate latency of 100ms for fault detection whereas SafetyNet [26] can tolerate 1 ms. Nakano *et al* proposed **Revive I/O** [221] is an extended version or variant of Revive. Both the approaches can manage output-commit problem (share validated checkpoint beyond recovery regions), generate global checkpoints and acquire data related to checkpoints in the memory. Doudalis *et al.* and Prvulovic *et al.* proposed **Euripus** [222] is a technique that has the ability to envisage redo-log and undo-log checkpoints, can capitalize on redo and undo logs thereby sidestep hardware replication which is a superfluous requirement. The rollback latency is enhanced by 30%, and performance overhead of 5% incurred in Euripus. Very high error rates are tolerated by Euripus that results in more than 95% system efficiency. Rishi *et al.* proposed **Rebound** [223] is a technique that constructs interaction sets between the check-pointing and rolling back using distributed algorithms, and cache coherence protocol to monitor inter-thread exchanges. It is a scalable recovery method suitable for many-core architectures.

**Research Challenges: -** In architectural core salvaging, a novel hardware-based thread migration exploiting cross-core redundancy such that cores are single instruction set architecture (ISA) compliant only. It may not be suitable for modern day heterogeneous multi-core system that supports multiple ISA or different ISA on cores. Concepts or proposals discussed above provide solutions to improve reliability of a homogeneous multi-core systems only. These concepts are not applicable to asymmetric multi-core systems with different ISAs. For example, Thread migration in homogeneous multi-core is less challenging as compared to multiple ISA heterogeneous multi-core system. Multi- ISA heterogeneous multi-core systems like AMD Fusion, ARM Tegra and Cell Broadband engine have demonstrated a strong budding alternative to homogeneous multi-core systems, in terms of improving performance, power efficiency and area efficiency. NVIDIA Tegra 4+1 Quad core has one low power battery saver core for low performance applications. Multi –ISA heterogeneous multi-core will become order of the day in near future. Challenges for providing resilient solutions with novelty for heterogeneous multi-core systems need to be addressed.

## 11 FAULT MODEL AND FAULT TOLERANT DESIGNS IN THROUGH SILICON VIAS BASED 3D NOC MANY-CORE SYSTEMS

The 3D IC designs are the ultimate alternatives for integrating hundreds of cores (say more-cores) with a smaller area on the chip [224]. The characteristics features which are top-notch in 3D IC design when compared to 2D planar IC design are performance enhancements [225], many-core packing onto die, and functionality. The significant edge in integrating diverse technologies [226] like GPGPU, MEMS, FPGA, DRAM, and analog onto a chip is a considerable advantage of the 3D IC model. The 3D IC designers face important difficulties in Through silicon vias (TSVs) due to increase in the vulnerability and defective rates of devices amalgamated with TSVs. Confirmable that node size of the transistor has been scaled down to physical limits, it cannot be reduced further thereby causing breakdowns and wear-outs [227,228]. At 8nm, the supply voltage has been reduced to 0.6v or beyond is expected to happen in the year 2018, thereby increases the susceptibility and more delicate to faults [229]. We examine (1) fault models (2) functional fault models (3) failure mechanism and failure models for TSV based NOC (4) fault tolerant 3D NOC routers and (5) digest on defect tolerance for TSV in this section.

**Fault Models for TSV based 3D Network on Chips: -** The conventional IC designs such as 2D IC designs have constraints which can be addressed by adopting decisive methods like 3D IC internally wire bonded by Through silicon vias (TSV). The routers in the 3-dimensional IC are connected by vertical links named TSVs placed along with the routers. The cross-bar switch, routing unit and input buffer are the important components of NOC router. The ICs capability and capacity can be expanded by integrating the devices vertically are integral to 3D IC designs. The 2D layers and TSVs hang together with micro bumps. The chip thickness is comparable to the TSV's maximum height of 200μm. 20μm is the diameter of vias in the TSV and may scale down to 5μm in near future [230]. The physical dimension of the TSV is briefly state here. It has 4μm-16μm via pitch, contact pads of 5μm-by-5μm size, diameter of vias of 2μm-8μm, 0.5μm oxide thickness, and 20μm-50μm layer thickness includes metallization and substrate [229]. The dimensions of TSVs such as pitches and diameter are larger in orders of two or three when compared to MOSFET gate lengths. The sequence of operations for integrating devices in the 3D ICs are (a) TSV fabrication involving chemical and mechanical process, (b)wafering, and (c) die bonding. Issues in 3D integration are reliability challenges and defects in wires, MOSFETs, micro-bumps and TSVs. CMOS fabrication and scaling rates are not similar to TSV fabrication. The **Figure 2.** [232] classifies the physical faults and logical fault models available. The issues in TSV are chip warpage, coupling and thermal stress. The defects are mapped to the logical fault models [232] in Table 14. Logic-level fault models in Table 14, denote corresponding physical faults. Classification helps in deriving formal equations in analytical method. Since physical level simulation takes longer period, classification provides a pragmatic view of the target model which comforts the construction of a simulation technique for signal observation and fault injection.

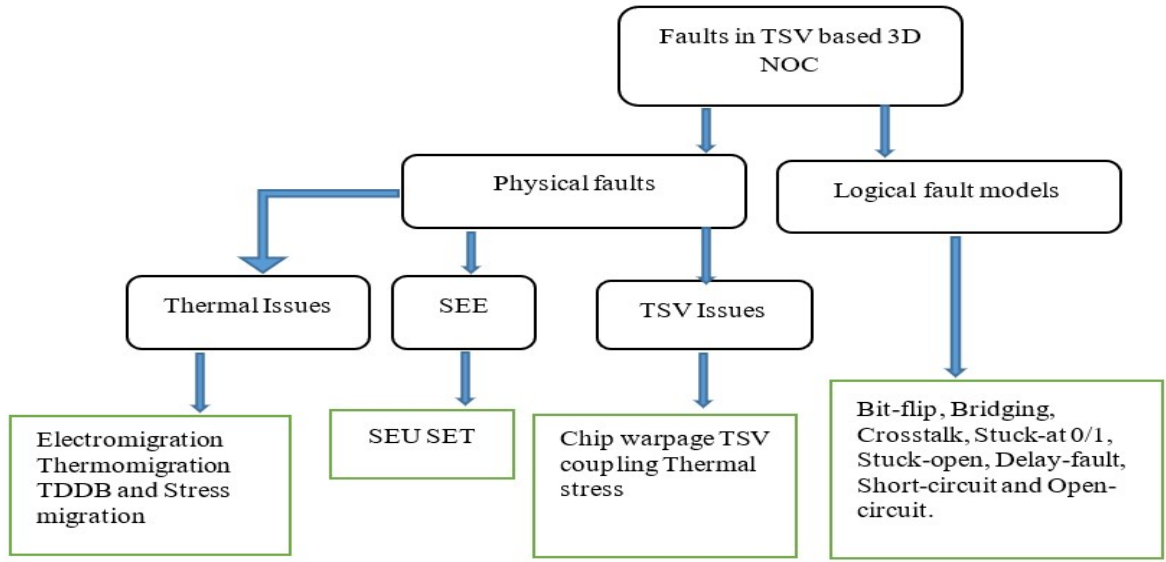

**Figure 2 Classification of physical faults and logical faults for TSV based 3D-NOC.**

Table 14 Mapping physical faults to corresponding logical fault models.

| Physical fault at TSV or causes (Thermal, SEE) | Failure Mechanism | Logical fault model |
|---|---|---|
| **SEE** | SEU | State transits in flip-flops or latches |
| **SEE** | SET | Stuck-at-0 or open |
| **TSV** | TSV coupling | Delay-fault <br> Bridging |
| **TSV** | Thermal stress | Delay-fault |
| **TSV** | Chip warpage | Open-circuit |
| **Thermal** | Electromigration | Delay-fault <br> Open-circuit |
| **Thermal** | Thermal cycling | Delay fault <br> Open-circuit |
| **Thermal** | Stress migration | Short-circuit <br> Open-circuit |
| **Thermal** | TDDB | Stuck-at-0-1 |

**Functional fault model for TSV based 3D NOC**: - The TSV based 3D NOC comprises of planar 2D component, links, NOC routers, inter and intra die connections and TSV. The sensitive elements vulnerable to faults are classified [232] and are shown in **Figure 3**. The effects of faults on sensitive elements cause malfunctioning or damage or loss in data. These functional damage or functional faults include Disconnection, Flit corruption, Timing jitter, Misrouting, Packet latency, Packet truncation, Packet drop and Header/data flit loss. The highlighted functional faults in **Figure 3** [232] are briefly described below.

**Packet latency**: - The occurrence of faults in the internal connections, TSV connections, and arbitration logic causes abnormal functioning in the routing logic thereby causes packet latency.

**Disconnection**: - The TSV body, micro-bump and contact may fail when they encounter fails. The disconnections or link open on the datapath occurs due to electromigration. The electromigration increases wire delay [149].

**Flit corruption**: - The fault occurrence in the TSV links, datapath of the router, crossbar switch, and intra-router links results in the corruptions of the packet or known as flit corruption.

**Timing Jitter:** - The provisional modifications in the signals from the declared places or slots in time is known as Timing Jitter. The variations in temperature and TSV coupling cause jitter.

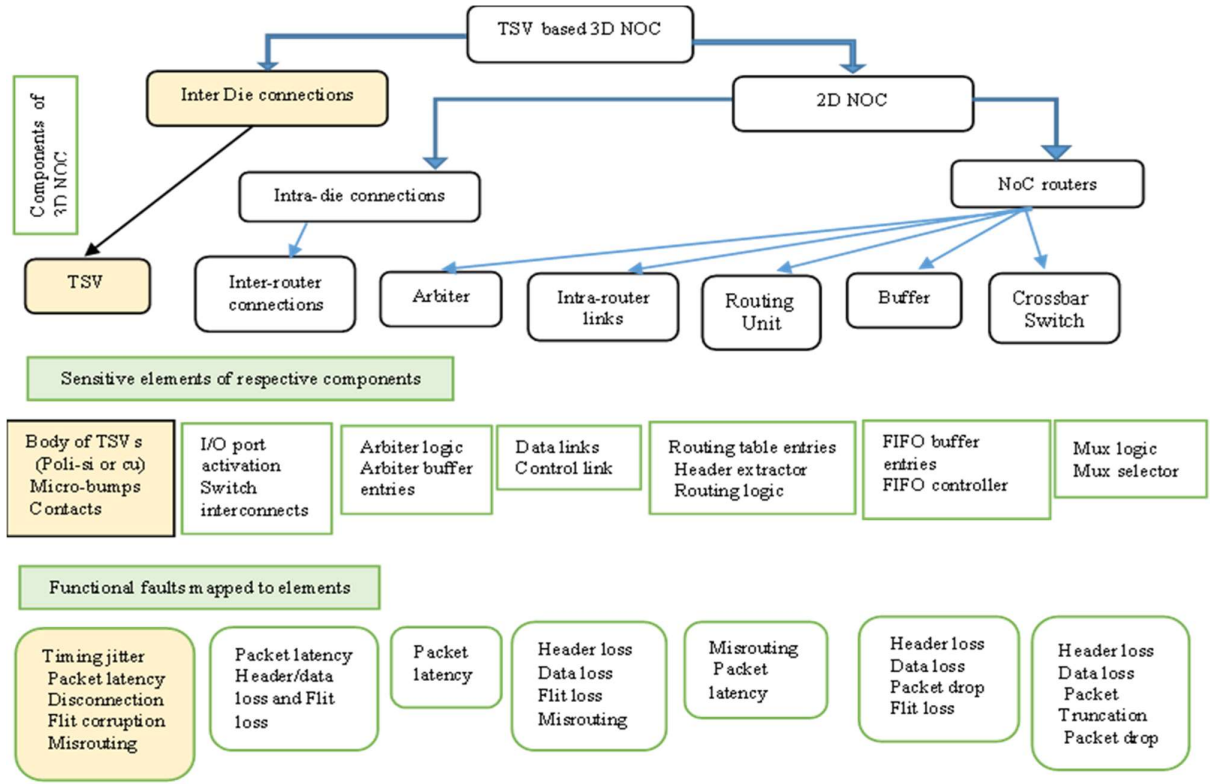

**Figure 3 Classification of functional fault models for TSV 3D NOC**

**Failure mechanism in TSV based 3D NOC:** - The failure mechanism of physical faults and their corresponding logical faults are listed in Table 1. The faults in the 3D NOC occur due to notable sources like SEE, thermal concerns and TSV issues [230,231,233]. The thermal stress, TSV coupling and chip warpage are the major causes of failures in the TSV [229,230]. **Chip warpage:** - The compression stress is caused due to uneven stretches or projections in the silicon and copper metal wires. The IC that experiences compression stress may suffer chip warpage. The vertical links or wires (TSV) arranged in an erratic manner would cause a unbalanced distribution of physical stress on the chip thereby enhancing the likelihood of physical damage. Attaching the TSV to the adjacent layer like wafer warpage [230] or placements of TSVs results TSV related defects. The annealing process can cause a fabrication defect otherwise known as wafer warpage. **TSV coupling** [232]**:** - The latency in the 3D signal path rises due to TSV capacitance. However, placement of buffer in between could reduce the latency overhead, as it preconditions extra power and area. The Miller effect introduce path delay in TSVs [234] [259] and may cause incorrect logic function thereby flips occur in the target signals. Very high delay is noticed in TSVs of smaller size with negligible overhead. TSV coupling at 8nm node size cause latency or coupling between adjoining TSVs [230,234]. The coupling can be inductive or capacitive in its effects. The magnetic field cause inductive coupling and electric field would result in the capacitive coupling in the TSVs region. In high frequency data transmission, inductive coupling is more rejective in behaviour [235]. **Thermal stress:** - In the early phase of TSV fabrication, the temperature is increased to perform electroplating and is decreased to room temperature. The outcome of this process results in the thermal stress between silicon and copper. The rapid changes in the thermal expansion results three levels of stress [236]. They are local stress, intrinsic stress, and extrinsic stress. In the intrinsic stress, micro-voids or pits are created in the copper due to stress, and pits or cracks grow in size in TSV and coalesce cause a delay fault or open circuit [261]. In extrinsic stress, different levels of thermal expansion fuelled by thermal cycling causes micro-bump distortion, cracks in the interfacial area among TSVs, TSV extrusion and deformity in metals cause open faults [236,237]. Local stress: The piezo resistivity may modify the deftness of the charge carriers [234] in the silicon substrate placed near TSV, thereby degrades performance [260]. This effect rises 10% latency in the every MOSFET could result in delay fault in the circuit [235]. SEE: The implications of SEE on TSV links is negligible. Other failure mechanism like Electromigration, HCI, NBTI, TDDB and thermal cycling do have significant impact on 2D IC and 3D IC designs.

**Failure Analysis for TSV coupling: -**

**TSV inductive coupling**: - MTTF is not a suitable parameter to analyse failure analysis of TSV coupling, since it is data dependent. The unanticipated total coupled voltage ($V_{Icoupltot}$) determines the coupling failure probability inductively induced in the TSV [232]. The inductive coupling voltage is denoted by β produced in TSVs. The αβ product is proportional to $V_{Icoupltot}$ on the loser TSVs. The configuration of active TSVs group and the current flow direction determines the value of α. The total coupled voltage $V_{Icoupltot}$ is equal to summation of the induced voltage on every assailant TSV, by applying faraday's law with presumption that effect of electromagnetic closeness is negligible, is given the Equation-(16). In the Equation-(16), the total coupled voltage is denoted $V_{Icoupltot}$, mutual inductance is denoted by $M_{v,i}$; total number of aggressors (or attacker) is denoted by N; current in the $i^{th}$ attacker TSV is denoted by $I_i$ ; coupling voltage on the $i^{th}$ attacker is denoted by $V_{Icoupl\ i}$. The mutual inductance is determined using the Equation-(17) [238], where length of the TSV is denoted l, distance between the $i^{th}$ aggressor from the victim TSV is denoted by d. Failure probability of TSV ($P_{find}$) resulted by inductive coupling is determined by the summation of occurrence frequency of α with respective failure probabilities is given in the Equation-(18). In the Equation-(18), α takes value 0,1,2,3, and 4, each denotes the flow of current and arrangement of TSV in 81 configurations.

$$V_{I\ coupl\ tot} = \sum_{i=1}^{N} V_{Icoupltot\ i} = \sum_{i=1}^{N} M_{v,i} \frac{dI_i}{dt} \approx \alpha\beta \quad (16)$$

$$M_{v,i} = \frac{\mu_0}{2\pi} [\ l \ln(\frac{l + \sqrt{d^2_i + l^2}}{d_i} + d_i + \sqrt{d^2_i + l^2})] \quad (17)$$

$$P_{find} = \frac{1}{81}[\ 19\ P_{0\beta} + 32P_{1\beta} + 20P_{2\beta} + 8P_{3\beta} + 2P_{4\beta}] \quad (18)$$

**TSV capacitive coupling: -** Symbiotic capacitance on the TSV deterministically brings down switching on a signal when an adjacent TSVs of the victim TSV execute transition in reverse direction. The charging and discharging of the victim TSV and adjacent TSVs due to miller effect [239] determines the values of the capacitive coupling. The charging and discharging of every TSV can be determined using the current flow direction between two dies, one die has the TSV driver, other has the load. In the Equation-(19), the total capacitance coupling noise is denoted by $TC_{tot}$ ; number of TSV aggressors is denoted by N. In the Equation-(20), $P_{fCap}$ denotes failure probability of TSV is determined by summing of the occurrence of every TSV factoring in all 234 patterns; occurrence probability is denoted by $P_{xc}$ , x can range between 0 to 8.

$$TC_{tot} = \sum_{i=1}^{N} |I_{vic} - I_{agg}| \quad (19)$$

$$P_{fCap} = \frac{1}{243}[\ 3\ P_{0C} + 16P_{1c} + 44P_{2c} + 64P_{3c} + 54P_{4c} + 32P_{5c} + 20P_{6c} + 8P_{7c} + 2P_{8c}] \quad (20)$$

**Failure Analysis for Thermal stress: -**

**Thermal Cycling: -** The thermal expansions in the different layers of 3D IC are not identical. It introduces interfacial crack in TSVs [151]. MTTF estimation in the stated in Equation-(21)[240]. In Equation-(21), ambient temperature is denoted by $T_{ambient}$ ; $T_{average}$ is the chip temperature on average basis; Coffin-Manson exponent constant is denoted by q; $A_{TC}$ is a empirical constant.

$$MTTF_{TC} = \frac{A_{TC}}{(T_{average} - T_{ambient})^q} \quad (21)$$

**Fault tolerant routers for 3D NOC: -** Ben Ahmed *et al.* proposed **3D-FTO Router** [241], it has a router architecture consisting of Traffic-Prediction-Unit (TPU), Random-Access-Buffer (RAB) mechanism, and Bypass-Link-on-Demand (BLoD), to provide fault tolerance at router level. RAB with TPU detects permanent faults, transient and intermittent faults in the input buffer and manages deadlock recovery. Traffic prediction unit collects the information in stipulated time interval about traffic load of each input port, using monitoring probes to find the best input port to receive the flits of faulty input buffer. Bypass-links are provided in the crossbar switches as escape channels when the faults increase. Fault control module enables the bypass link when there is a fault and prevents flits requesting faulty cross link. The Look-Ahead-Fault-Tolerant routing algorithm performs minimal routing and selects the direction with highest next hop diversity. 3D-FTO continues to perform with 250 faults dispersed along crossbar links, buffer-slots, and links. Dang et al. proposed **3D-FETO SYSTEM** [242]. It has backbone component called adaptive SHER -3D router. Permanent fault detection and recovery is similar to 3D-FTO router. It uses pipeline computation redundancy method to soft errors. It performs soft error detection and recovery in next port computation (NPC) and switch allocation (SA) module by comparing the results from original NPC, SA with redundant NPC and SA. It is observed that SHER-3D router has 1.9x and 1.49x MTTF improvement for permanent faults and soft errors respectively over baseline router. SHER-3DR's hard fault tolerance is 2.96 times better the baseline 3D-OASIS-NOC router [243,244].

Poluri et al. proposed **SHEILD** [245], an improved router design where router supplements an extra routing computation unit in every input-port to guarantee resilience in the route computation stage. Two-stage separable virtual channel allocator for performing virtual channel allocation is used. The article presents alternate path to circumvent the faulty arbiters in the Switch allocation stage. In the crossbar traversal stage, two paths are provided that leads to a specific output-port of the crossbar. SHEILD tolerate both permanent faults and soft errors. In this article, new metric Soft Error Improvement Factor has been introduced and SHEILD is three times more soft error tolerant than the baseline unprotected router with respect to the new metric. SHEILD has 6x more MTTF than the baseline router.

Liu et al. proposed **Data-path salvaging** [246] the data path (like links, input buffers and crossbar) is sliced by splitting the path into smaller horizontal paths. This method avoids redundancy. The one of the slices fails, other slice can be used using time division multiplexing mode. Silicon protection factor (SPF) is defined as the router before becoming non-operational, number of faults it can tolerate is normalized by the area penalty of the method The SPF of Data-path salvaging scales from 5.00 to 13.07. Hossein Zadeh et al. [247] proposed redundant router and link bypassing methods for application specific NOC with huge hardware cost. There are other router architectures that restrict the usage of TSVs in router to reduce risk of defective TSV, thereby improving the reliability of the structure. Limited usage of TSVs is an impediment to upcoming exploration of high bandwidth applications deployed on TSV based NOC.

**Defect tolerance methods for TSV in 3D NOC: -**

**TSV Redundancy: -** Two redundant TSVs are provided with four signal TSV in a TSV bundle in Samsung memory [248]. It results in high redundancy ratio of 1:2. A signal shifting is done with chain linked TSVs [249] with one spare used for connecting chain of multiplexors. Signals transmitted shifted from one defective TSV to other good TSVs is a form repairment in the TSVs-chain. Here, redundancy ratio is variant factor and depends upon the length of TSV. Loi et al. proposed usage of TSV grids as interconnect links and augmented the boundary regions of TSV with additional or extra TSVs [250]. Extra TSVs in the same row/column of the crossbar is linked with signal TSVs and can bear one failure in TSV grid. For a N x N TSV grid, 1: N redundancy ratio is used.

**TSV clusters: -** All the redundancy approaches replace faulty TSVs with the adjacent or neighbouring TSVs. The chip warpage, capacitive and inductive TSV coupling and thermal stress cause cracks in micro-bumps and TSVs. Hence redundancy-based repair approaches may be less effective. An innovative repair path routing approach to search for a redundant TSV that not in neighbourhood is proposed by [262]. Demerit of this approach is to meet timing constraint and length of the interconnects on the alternate repair path. The disaster by clustered faults in TSV clusters can be recovered [251] by constructing virtual TSV grid where topological neighbour need not be mapped to physical neighbour. Cluster defect tolerance approaches makes TSV based 3D NOC system highly reliable.

**Timing redundancy: -** The signals that pass through the TSV links have non-rigid timing constraints to provide defect tolerance. The data released form faulty TSVs are serialized [252] and reoriented towards fault-free TSVs. The self-regulated de-multiplexer and multiplexer is present in every serial channel restricted to vertical channels attaining serialization and de-serialization asynchronously. Kologeski et al. protracted the idea of link serialization [253] for optimal use of partial open circuit / resistive TSVs to send data bits using them at low speed when compared to fault-free TSVs. Usage of low quality TSVs at slower rate of transmission compensates the delay variations of TSVs. On field swapping of TSV-faulty for a signal with rigid timing constraint with another signal with non-rigid timing constraint is proposed in TSV- repair approach [254]. Online mechanism for delay fault testing and respective on-field repair algorithms are part of the solution provided in TSV-repair.

## 12 CORE LEVEL FAULT TOLERNACE METHODS FOR MANY-CORE SYSTEMS

For Network-on-chip based many-core systems, core-level redundancy approaches are the most suitable approach for permanent fault recovery. Few articles discuss self-organization of cores in field by using diagnosing methods. In this approach appropriate communication channels are established with fault free cores. Scalable distributed approaches use software-based BIST locally on each core. Reconfiguration based recovery procedure are executed and make many-core system operate in presence of faults. We examine only hardware-based solution for detecting, diagnosing and reconfiguration-based recovery addressing permanent faults for many cores system.

**Core level redundancy: -** Zhang *et al.* proposed X+S defect tolerance [255] is a technique that show improvements in yield. This method uses X number of cores fabricated with S spares cores. X number of cores will be always available to customers in presence of faulty cores. If there are at most S number of faulty cores, then only chip can be repaired. In this method, in field reconfiguration is not supported and does not consider faults in NOC. The spare cores are not operational during normal course. Core level redundancy strategy has huge hardware overhead cost (>100%).

**Core self-organization: -** A self-core organization approach [256,257] proposed by Zajac *et al.* and Collet *et al.* to support hard error recovery in many-core systems. Each core executes diagnosis program locally stored (in like flash memory) and self-testing is performed. The good core detects fault in one of its neighbours, it aborts all data transmission with the same. To examine point-point communication and routers, hardware-based method are used in NOC diagnosis. In finding a communication route, all nodes broadcast message to all so that it reaches the other good nodes. Special hardware mechanism is used by routers to broadcast message and append local routing decision in the header of the packet. Recipient node receives this packet, and sends an acknowledgement to the sender which has adopted the same route. Lastly, after all the acknowledgments are received, sender collects all the routes to contact other good nodes.

**Distributed local diagnosis in cores**: - Scalable in-field test architecture [258] proposed by Kamran *et al.* and Navabi *et al.* dispense test codes among homogeneous cores in many-core processors. SBST are executed in every core and self-testing is performed. Smaller sized test-codes of SBST are stored in a chip memory or off-chip memory. A specific hardware procedure is developed to broadcast the test-codes stored in the buffers of a cluster (or group of cores in region). The test-codes are locally executed in the cores. As soon as cores complete the execution of the test-codes, additional hardware device collects the results of execution and consolidates them to analyses the cores to identify faults. Additional requirement for this approach is the decentralized recovery methods that permit the cores to re-organize themselves and function normally in existence of faults. It has permanent fault coverage of 93%.

**Research challenges in TSV based 3D NOC many-core systems: -**

(a) Already, 2D NOC are used in FPGA based accelerators and GPGPUs to address the on-chip bandwidth issue. 3D integration of FPGA and GPGPU is next natural choice for enhancements in performance and lower cost. Such 3D integration result in new fault models, failure mechanism and new mitigation techniques are needed to improve the reliability.

(b) Increase in the performance largely due to increase in density of TSVs in the die. Higher density of TSVs will lead to capacitive and inductive coupling between TSVs. Unexpected parasitic signals cause capacitive coupling. Mitigation techniques for capacitive and inductive coupling are not well studied. Complexity of modelling capacitive and inductive coupling increases and it involves electromagnetic analysis of coupling with equivalent electrical model for TSVs. In addition, design of routers needs overhaul and routing algorithms needs to be redesigned.

(c) Integrating modern non-volatile memories into TSV based 3D NOC will introduce new faults models, failure mechanism and thereby new fault tolerant techniques for these systems.

(d) Heterogeneous many-core systems at the architecture level are the future high performance computing system. The new design alternatives at the core level are suggested for further exploration. They are (a) Multiple ISA heterogeneous many-core system with fault tolerant instruction emulating cores or Dependable Instruction Emulating Cores (DIMC). We may call it as TSV based 3D many cores with DIMC. (b) Multiple ISA heterogeneous many-core system with Energy Efficient fault tolerant or Dependable Cores (EEDC), and (c) In TSV based 3D NOC many-core system, core performing coarse grain redundant re-execution at SRAM based FPGA layer. It is like leading core at L1 layer with trailing core or processing element at (FPGA) L2 layer.

**CONCLUSION**

In this article, destructive effects and non-destructive effects of radiation on Digital CMOS and SRAM based FPGA are described in great detail. Understanding of these effects are essential for validating the logical fault models and failure models to intuitively analyse fault mitigation techniques at logic, micro-architectural and architectural levels. Gamut of fault mitigations methods focussing on single event upsets, timing faults and permanent faults are described quantitatively stating improvements in reliability at the respective level of abstraction. Significance of NOC, 3D integration and TSV based 3D NOC are presented with relevant fault models and failure mechanism. Prominent resilient routers are only examined. In the field defect tolerance method and diagnosis methods for many-core systems are highlighted with fault coverage for permanent faults. New design alternatives at architecture level are presented to further explore research opportunities and investigate mapping of application to each one of the designs by specifying data availability and integrity requirements.